\documentclass[prd,showpacs,groupedaddress,superscriptaddress,twocolumn,
10pt,nofootinbib,floatfix,preprintnumbers,letterpaper]{revtex4}
\usepackage{hyperref,amssymb,amsmath,graphicx,xcolor,cancel}

\newcommand{\lsim}{\raisebox{-0.7ex}{$\stackrel{\textstyle <}{\sim}$ }}

\begin{document}

\begin{figure}[!t]
\vskip -1.1cm
\leftline{
\includegraphics[width=3.0 cm]{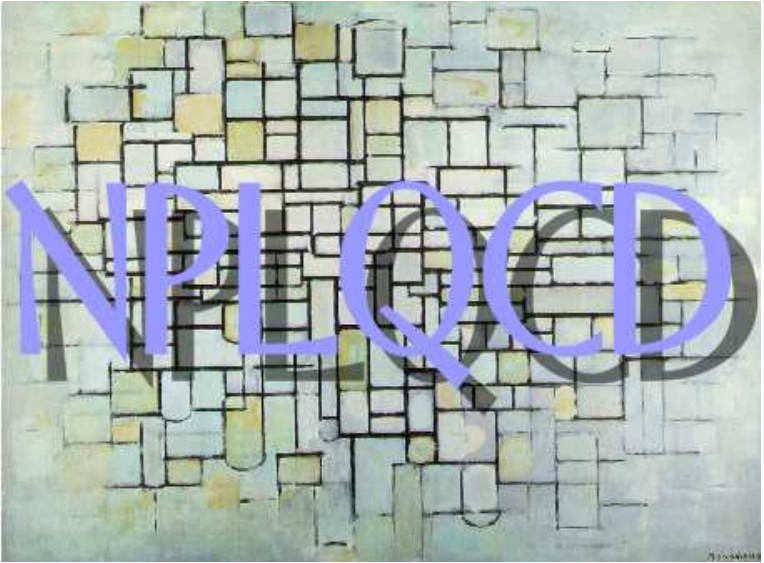}}
\vskip -0.5cm
\end{figure}

\title{Quarkonium-Nucleus Bound States from Lattice QCD}

\author{S.R.~Beane} 
\affiliation{Department of Physics, University of Washington, Seattle, WA 98195-1560, USA}
  
\author{E.~Chang}
\affiliation{Department of Physics, University of Washington, Seattle, WA 98195-1560, USA}
\affiliation{Institute for Nuclear Theory, University of Washington, Seattle, WA 98195-1560, USA}

 \author{S. D. Cohen}
\affiliation{Institute for Nuclear Theory, University of Washington, Seattle, WA 98195-1560, USA}

\author{W. Detmold} \affiliation{
Center for Theoretical Physics, Massachusetts Institute of Technology, Cambridge, MA 02139, USA}

\author{H.-W. Lin}
\affiliation{Department of Physics, University of Washington, Seattle, WA 98195-1560, USA}

\author{K.~Orginos}
\affiliation{Department of Physics, College of William and Mary, Williamsburg,
  VA 23187-8795, USA}
\affiliation{Jefferson Laboratory, 12000 Jefferson Avenue, 
Newport News, VA 23606, USA}

\author{A.~Parre\~no}
\affiliation{Dept. d'Estructura i Constituents de la Mat\`eria. 
Institut de Ci\`encies del Cosmos (ICC),
Universitat de Barcelona, Mart\'{\i} Franqu\`es 1, E08028-Spain}

\author{M. J. Savage}
\affiliation{Institute for Nuclear Theory, University of Washington, Seattle, WA 98195-1560, USA}

\collaboration{NPLQCD Collaboration}

\date{\today}

\preprint{INT-PUB-14-049}
\preprint{NT@UW-14-23}
\preprint{MIT-CTP-4600}


\begin{abstract}
  Quarkonium-nucleus systems are composed of two interacting hadronic
  states without common valence quarks, which interact primarily
  through multi-gluon exchanges, realizing a color van der Waals
  force.  We present lattice QCD calculations of the interactions of
  strange and charm quarkonia with light nuclei.  Both the
  strangeonium-nucleus and charmonium-nucleus systems are found to be
  relatively deeply bound when the masses of the three light quarks
  are set equal to that of the physical strange quark.  Extrapolation of
  these results to the physical light-quark masses suggests that
  the binding energy of charmonium to nuclear matter is 
  $B^{\rm NM}_{\rm phys} \lsim 40~{\rm MeV}$.
\end{abstract}

\maketitle

\section{Introduction}
\label{sec:intro}

Since quantum chromodynamics (QCD) was first proposed as the underlying theory of the strong interactions, 
enormous progress has been made in the understanding of hadrons as composite objects formed from quarks and gluons.
A particularly interesting consequence of the extended nature of hadrons is their 
susceptibility to chromo-polarization, which allows for hadronic interactions that are distinct from 
meson-exchanges  which dominate the long-range
forces between nucleons.
The effects of color polarization can be isolated and explored by considering
hadronic systems without shared valence quarks, thereby eliminating the possibility of quark-exchange interactions and Pauli blocking.

The significance of a color van der Waals force (so called by analogy to the electromagnetic effect) was first appreciated by 
Brodsky, Schmidt and de~Teramond (BSdT) in 1990~\cite{Brodsky:1989jd}. 
They observed that the rapid variation in the spin-spin correlation in $pp$ scattering at a scattering angle of $\theta=90^{\circ}$ 
near the open charm production threshold ($\sqrt{s} \sim 5~{\rm GeV}$)
may be indicative of a strong attractive interaction between charmonium and the di-proton system.
In terms of quarks and gluons,  these systems interact through multi-gluon exchanges,
which  manifest themselves as two-pion exchange interactions at long distances, but 
which are not expected to generate repulsion at short distances.
Using a  Yukawa toy model to describe the charmonium-nucleus interactions, BSdT predicted bound states for nuclei with atomic numbers
$A\ge 3$, with binding energies of
$B_{^3{\rm He}\, \eta_c } = 19\text{ MeV}$, $B_{^4{\rm He}\, \eta_c} = 140\text{ MeV}$, 
and as deep as $B_{^9{\rm Be}\,\eta_c} =  407~{\rm MeV}$.
Subsequent works have refined these calculations, 
starting with the observation by Wasson~\cite{Wasson:1991fb} that the extended volume of large nuclei must modify the  form of the 
potential, which had been assumed to scale with $A$ in Ref.~\cite{Brodsky:1989jd}. 
This more realistic model suppresses the binding energies 
compared with those obtained in Ref.~\cite{Brodsky:1989jd}, leading to estimates of 
$B_{^3{\rm He}\, \eta_c} = 0.8\text{ MeV}$, $B_{^4{\rm He}\, \eta_c} = 5\text{ MeV}$, 
and  which rapidly saturate to  $B^{\rm NM} \lsim 30~{\rm MeV}$ in nuclear matter (NM). 
The heavy-quark expansion, in which the binding energies have expansions 
in inverse powers of the heavy-quark mass, $M_Q$, 
and in the radius of the quarkonium, $r_{\overline{Q}Q}$, 
was applied to these systems in Ref.~\cite{Luke:1992tm}.
Using an operator product expansion,
the dominant effects arise from 
matching to the leading  dimension-seven operators involving the quarkonium and two gluons
with coefficients that scale as $r_{\overline{Q}Q}^3$.
At NM density,  a binding of $\sim 10$~MeV was found for the $J/\psi$. 
However, since the chromo-polarizability  depends upon the radius of the charmonium, the excited state $\psi'$, 
which is loosely bound and has large radius (about 1.8~fm), may 
be more
deeply bound to nuclei, although the  techniques used for that analysis become 
unreliable for these larger systems.
Nonperturbative modifications to the interactions and nuclear binding of quarkonia have been explored through the inclusion of hadronic-exchange effects, 
e.g. Ref.~\cite{Brodsky:1997gh}.
A summary of the predictions for charmonium binding to the lightest nuclei and NM is given in Table~\ref{tab:phenpred}.
\begin{table}[!t]
\begin{tabular}{|c|ccc|cc|}
\hline
 & \multicolumn{3}{c|}{\  Binding Energy (MeV)\  } & \multicolumn{2}{c|}{\ Binding Energy (MeV) } \\
Ref. &\  $^3$He $\eta_c$ \ &\  $^4$He $\eta_c$ \ &\  NM $\eta_c$
     & \   $^4$He $J/\psi$ \  &\  NM $J/\psi$  \\
\hline
\cite{Brodsky:1989jd}
 &  19  &  140 &  
  &     &      \\
\cite{Wasson:1991fb}
  &  0.8 &    5 &   27 
   &     &      \\
\cite{Luke:1992tm}
   &     &     &  10    
   &     &   10
 \\
\cite{deTeramond:1997ny}
  &  $\ast$ & $\ast$ &    9 
    &     &      \\
\cite{Ko:2000jx}
  &     &     &     
   &     &    5 
\\
\cite{Tsushima:2011kh}
   &     &     &     
    &    5 &   18 
    \\
    \cite{Yokota:2013sfa} 
    & & & 
    & 15.7 &
\\
\hline
\end{tabular}
\caption{
Estimates for the binding energies of charmonium to light nuclei
and nuclear matter  (in MeV) from selected models.
A ``$\ast$'' indicates the system is predicted to be unbound, while blank entries indicate that the system was not addressed.
}
\label{tab:phenpred}
\end{table}

In addition to charmonium interactions with nuclei, 
the interactions of  bottomonium and  strange-quarkonium with nuclei have also been considered. 
The heavy-quark expansion works well for bottom quarks~\cite{Luke:1992tm}, 
from which it is found that, because of  its smaller radius, 
bottomonium is less bound to NM than charmonium, with an estimated binding energy of $\sim 4~{\rm MeV}$. 
Strange quarkonia binding to nuclei has also been considered previously, and in particular,
the $\phi$ has been predicted to 
have a binding energy of $\sim 40~{\rm MeV}$ to NM~\cite{YamagataSekihara:2010rb}.
Although the strange pseudoscalar, $\eta_s$, 
mixes strongly with the light-quark pseudoscalars to form the physical $\eta$ and $\eta^\prime$, for theoretical purposes it can be treated as a pure $\bar s s$ state, in a manner analogous to the  $\eta_c$.
The work of Ref.~\cite{Haider:1986sa} finds 
that the $\eta_s$ does not bind to nuclei with $A<12$, 
but does bind to NM with $B^{\rm NM} \sim 17~{\rm MeV}$, 
while Ref.~\cite{Tsushima:2011kh} finds  
the $\eta_s$ binds to NM with $B^{\rm NM} \sim 90~{\rm MeV}$.

Despite the general agreement among theorists that charmonium-nucleus bound states should exist,  
the predictions for the binding energies are quite disparate, and such  systems remain to be discovered experimentally despite many attempts to produce them. 
The latest experimental programs in this area include ATHENNA~\cite{ATHENNA} as part of the 12-GeV program at Jefferson Lab, PANDA at FAIR~\cite{fair,Barabanov:2013cna} and
efforts at J-PARC~\cite{jparc}. 
A signal of $^3\text{He}$ $\eta$  was reported by MAMI~\cite{Pfeiffer:2003zd} with $B_{^3{\rm He}\, \eta} \sim 4~{\rm MeV}$, 
but the result could not be confirmed by COSY-GEM which, however, did report evidence for  a bound  $^{25}\text{Mg}$ $\eta$ system
with $B_{^{25}{\rm Mg}\, \eta} \sim 12~{\rm MeV}$~\cite{Budzanowski:2008fr}.

In order to guide the present and future experimental programs aiming to discover and explore quarkonium-nucleus bound states, 
it is important to perform QCD calculations of these systems. 
Lattice QCD (LQCD) is currently the only reliable technique for  such calculations in the nonperturbative regime, 
and exciting progress has been made in recent years applying LQCD to light nuclei~\cite{Beane:2010hg,Beane:2011iw,Yamazaki:2011nd,Yamazaki:2012hi,Beane:2012vq,Beane:2013br,Beane:2014ora}. 
In addition, an early calculation of the color polarizabilities of mesons was performed~\cite{Detmold:2008bw}, in which it was found that Bose gases of pions or kaons
become color-polarized when  in the presence of static color sources. 
This has been extended to the case of 
charmonium and bottomonium  interactions with many pion systems~\cite{Detmold:2012pi}.
Lattice QCD calculations of the scattering of quarkonia and single 
nucleons have been previously performed~\cite{Yokokawa:2006td,Liu:2008rza,Kawanai:2010cq,Kawanai:2010ru}.
Quenched calculations reveal a negative scattering length (with the nuclear physics convention), resulting from an attractive interaction, 
but the results are consistent with a volume-independent negative energy shift, as would arise 
from a bound state.
Calculations with $n_f=2+1$~\cite{Kawanai:2010ru} at a pion mass of $M_\pi\sim 640~{\rm MeV}$ 
yield a relatively small and negative scattering length,  a large effective range, but not a bound state.
The HAL QCD modeling method has been used to extract interpolating-operator- and energy-dependent quarkonium--light-hadron potentials, e.g. Ref.~\cite{Kawanai:2011jt}.

In this work, we demonstrate the existence of quarkonium-nucleus bound states 
for $A<5$, and calculate their binding energies, at the flavor SU(3)-symmetric point with 
unphysical values of the light-quark masses corresponding to that of the physical strange quark, 
resulting in a pion of mass $M_\pi\sim 805~{\rm MeV}$.
The same lattice technology and parameters, with the addition of the charmed quark, are used 
as in the calculations of light nuclei presented in Refs.~\cite{Beane:2012vq,Beane:2013br,Beane:2014ora}.
While   calculations are performed in multiple lattice volumes, only one lattice spacing has been employed.

In Sec.~\ref{sec:setup}, the lattice QCD calculations performed in this work are described. 
The methods used to analyze the correlation functions, and the binding energies extracted from them, are  presented in Sec.~\ref{sec:results}. 
Boosted systems are found to present some unexpected challenges, 
which are discussed in Sec.~\ref{sec:boosted}. 
Finally, we present our conclusions and discuss the future  lattice QCD prospects for quarkonium-nucleus systems in Sec.~\ref{sec:end}.

\section{Lattice QCD Methodology}
\label{sec:setup}

Three ensembles of  gauge-field configurations at the SU(3)-flavor symmetric point, 
where  $M_\pi=M_K \sim 805$~MeV, at a single lattice spacing of $b = 0.145(2)~{\rm fm}$ (determined at this unphysical mass) were used in this work. 
The L\"{u}scher-Weisz gauge \cite{Luscher:1984xn} action was used with a clover-improved quark action \cite{Sheikholeslami:1985ij} with one level of stout smearing
($\rho=0.125$)~\cite{Morningstar:2003gk}. 
The clover coefficient was set equal to its tree-level tadpole-improved value, $c_\text{SW}=1.2493$, 
a value that is consistent with an independent numerical study of the nonperturbative 
$c_\text{SW}$ in Schr\"{o}dinger functional scheme \cite{EdwardsPC}.
The ensembles have spatial extent $L\sim 3.4, 4.5$ and $6.7~{\rm fm}$, and each consists of $O(10^4)$ 
evolution trajectories.
Large volumes are necessary for the study of bound states in lattice QCD even at heavy quark masses, and we have previously published results for the spectroscopy of light nuclei and hypernuclei~\cite{Beane:2012vq}, 
and for nucleon-nucleon scattering properties~\cite{Beane:2013br}, 
obtained from them. 
The relevant features of these ensembles 
are given in Table~\ref{tab:param} (further details can be found in Refs.~\cite{Beane:2012vq,Beane:2013br}).
Somewhat fewer measurements are used in the present work than in Refs.~\cite{Beane:2012vq,Beane:2013br}.
\begin{table}[!t]
\begin{tabular}{|c|cc|ccc|}
\hline
$L^3\times T$ & $N_\text{cfg}$ & $N_\text{src}$ & $aM_\pi$  & $M_\pi L$ &  $aM_N$  \\
\hline
$24^3\times 48$ & 1894 & 96 & 0.59388(14) & 14.3 & 1.2042(5)   \\\hline
$32^3\times 48$ & 3093 & 48 & 0.59451(08) & 19.0 & 1.2046(8)   \\\hline
$48^3\times 64$ &  614 & 64 & 0.59446(11) & 28.5 & 1.2047(9)   \\\hline
\end{tabular}
\caption{
Details of the ensembles of  gauge-field configurations used in the present calculations, 
including the lattice dimensions, 
number of configurations per ensemble, $N_\text{cfg}$, 
number of sources used per configuration $N_\text{src}$,
along with the pion and nucleon masses.
[Note that as this involves only a subset of the number of sources used in our calculations of nuclear binding energies and nucleon-nucleon
scattering~\cite{Beane:2012vq,Beane:2013br}, the light-hadron masses in this table have somewhat larger uncertainties.].
}
\label{tab:param}
\end{table}

Multiple different correlation functions for the strangeonium- and charmonium-nucleus
systems were calculated on the 
 ensembles of lattice gauge-field configurations described above. 
The correlation functions of these systems are simply 
the product of the individual correlators of the component subsystems on each gauge field for a given source location. 
Consequently, the
nuclear correlation functions previously calculated were re-used, 
and additional computational resources  were only expended on the quarkonium correlation functions.
The nuclear correlation functions were  produced  using the recursive algorithm of Ref.~\cite{Detmold:2012eu}, 
and a detailed study and results for nuclear bindings and interactions can be found in Refs.~\cite{Beane:2012vq,Beane:2013br}. 
In the current study, we   focus on  the 
nucleon $(J^\pi\!=\!\tfrac{1}{2}^+)$, 
deuteron $(J^\pi\!=\!1^+)$, di-neutron $(J^\pi\!=\!0^+)$,
$^3\text{He}$ $(J^\pi\!=\!\tfrac{1}{2}^+)$ and 
$^4\text{He}$ $(J^\pi\!=\!0^+)$.
We have  previously calculated correlation functions of the strange mesons, $\eta_s$ and $\phi$, 
 for a range of momenta on the same ensembles.
To calculate charmonium correlation functions, 
charm-quark propagators were produced using the relativistic heavy-quark (RHQ) action~\cite{ElKhadra:1996mp}:
 \begin{multline}
S_Q=\sum_{x,x'} \overline{Q}_x \Big(m_0 + \gamma_0 D_0 - \frac{a}{2} D_0^2 
+\nu\left(\gamma_i D_i-\frac{a}{2} D_i^2\right) \\
-\frac{a}{4}c_\text{B}\sigma_{ij} G_{ij} 
-\frac{a}{2}c_\text{E}\sigma_{0i} G_{0i}\Big)_{xx'} Q_{x'},
\end{multline}
where $Q_x$ is the heavy-quark field at the site $x$, 
$\gamma_\mu$ are the Hermitian Dirac matrices,
$\sigma_{\mu\nu}$ is defined through $i \left[\gamma_\mu,\gamma_\nu\right]/2$, 
$D_\mu$ is the first-order lattice derivative,
and $G_{\mu\lambda} = \sum\limits_a T^a G_{\mu\lambda}^a$ is the Yang-Mills field-strength tensor.
The coefficients $\nu=1.295$ 
and $m_0=0.1460$ 
were tuned to recover the spin-averaged $\eta_c$ and $J/\psi$ experimental masses and  low-energy dispersion relations,
while
$c_\text{E,B}$ were set to their tree-level tadpole-improved values, 
$c_\text{B}= c_\text{SW} \nu = 2.24363524134292$ 
and $c_\text{E} = c_\text{SW} (1+\nu)/2 = 1.9880860536224$.
(For a more detailed discussion of this tuning, see Ref.~\cite{Brown:2014ena} and references therein.)
\begin{figure}
\includegraphics[width=0.45\textwidth]{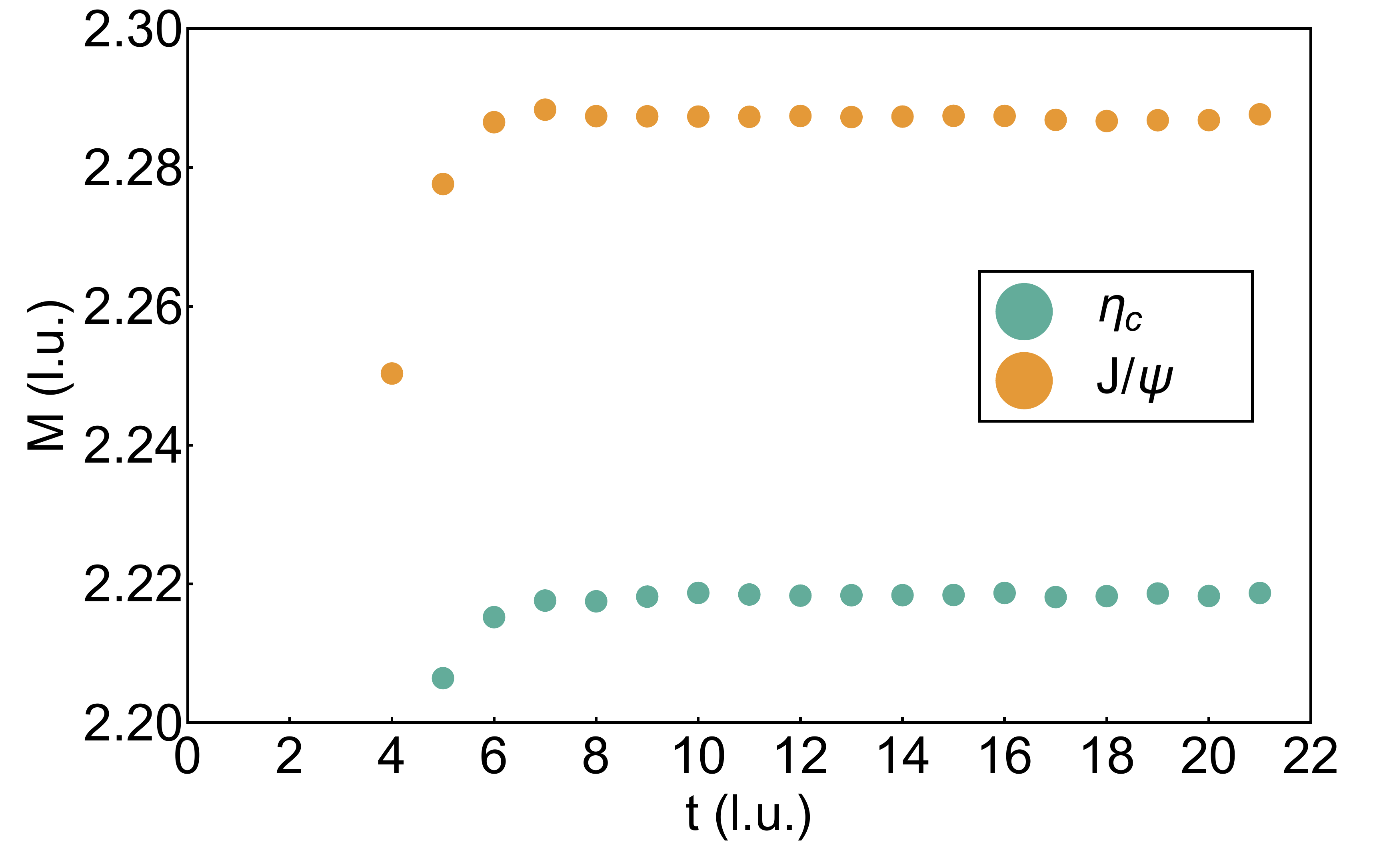}
\caption{
The effective mass plots associated with the $\eta_c$ and $J/\psi$ formed from linear combinations of the
smeared-point and smeared-smeared correlation functions.
}
\label{fig:ccemps}
\end{figure}
Analysis of the correlation functions, 
that give rise to the effective mass plots shown in Figure~\ref{fig:ccemps},
including all statistical and systematic uncertainties, gives
masses  ($M_i(L)$) of
$M_{\eta_c}(24)=3012(33)~{\rm MeV}$,
$M_{\eta_c}(32)=3012(33)~{\rm MeV}$,
$M_{J/\psi}(24)=3105(34)~{\rm MeV}$ and
$M_{J/\psi}(32)=3106(34)~{\rm MeV}$,
and mass splittings ($\Delta M_i(L)$) of
$\Delta M(24) = 93(1)~{\rm MeV}$ and 
$\Delta M(32) = 93(1)~{\rm MeV}$,
where the dominant uncertainty is that from the lattice spacing.
Table~\ref{tab:c} shows the ``speed of light'' for each hadron obtained from
quadratic fits to the squared-energy versus squared-momentum for the chosen RHQ parameters.
\begin{figure}
\includegraphics[width=0.45\textwidth]{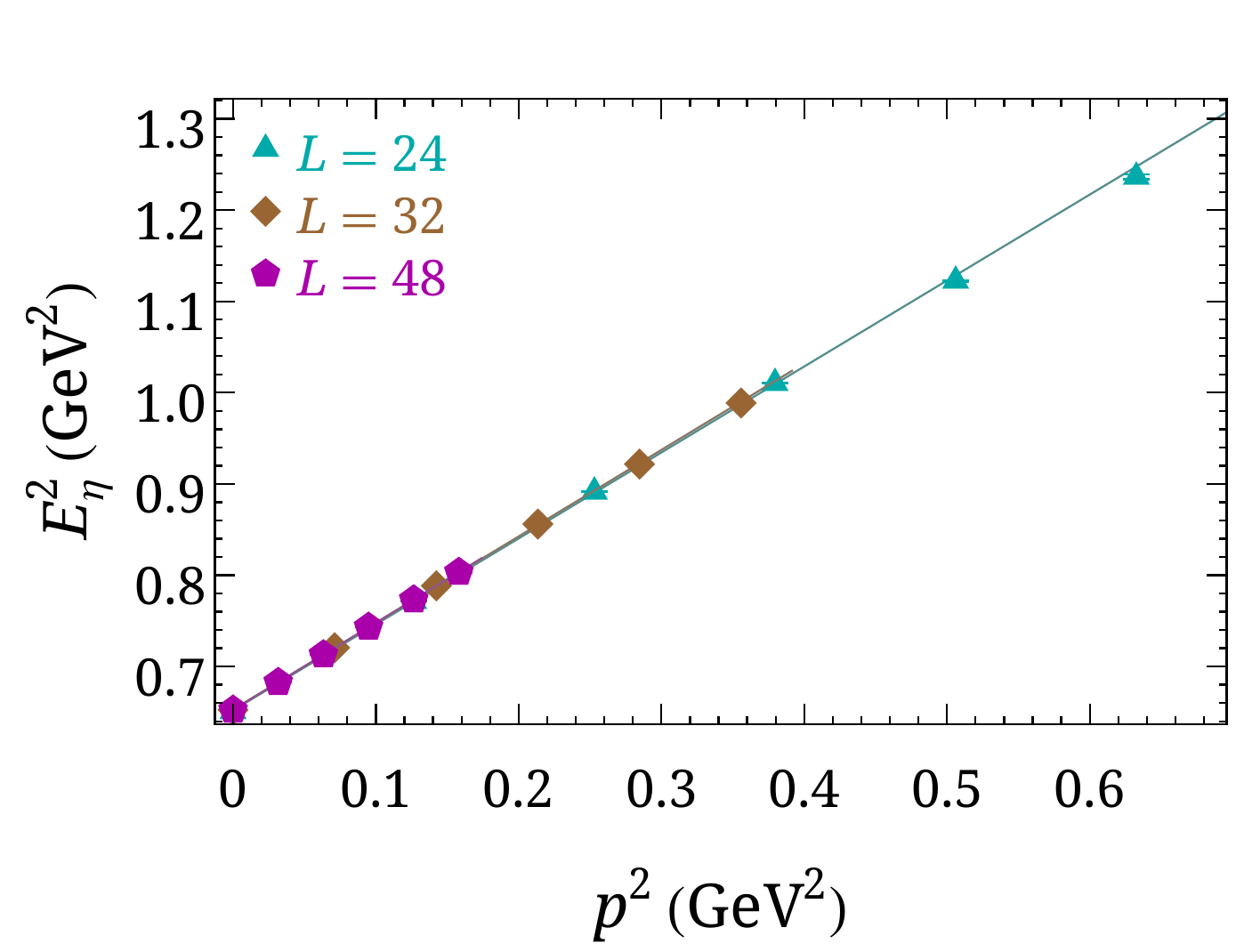}
\includegraphics[width=0.45\textwidth]{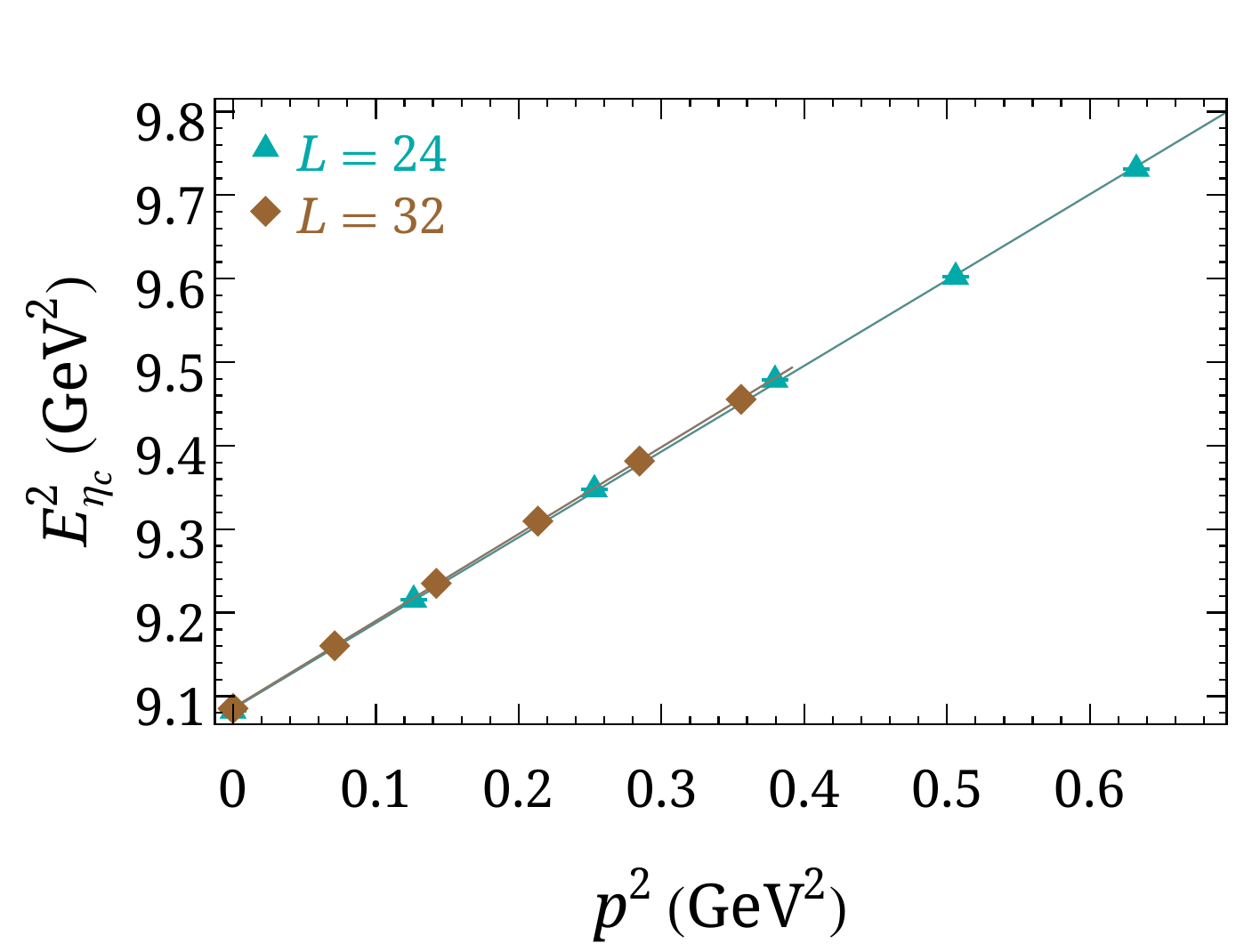}
\caption{
The dispersion relations of the  $\eta_s$ and $\eta_c$. 
The blue triangles,
brown diamonds and purple pentagons show results from the  $L=24,32$ and $48$ ensembles, respectively.
The curves correspond to linear fits to the $p^2\le 0.4~{\rm GeV}^2$, and show small 
quadratic contributions at higher $p^2$.
}
\label{fig:disp}
\end{figure}
Figure~\ref{fig:disp} shows the calculated dispersion relations for the $\eta_s$ and $\eta_c$, ~\footnote{
Unfortunately, charmonium correlation functions were not calculated in the $L=48$ volume.
This was a consequence of this work occurring after the production of the nuclear correlation functions, and
practical aspects associated therewith.  
The same is true for the $N$-$\eta_s$ correlation functions in this volume.
}
which are representative of the dispersion relations for the quarkonia considered in this work,
and demonstrate that the ${\cal O}(a m)$ effects in charmonium are well controlled.
\begin{table}
\begin{tabular}{|c|cc|cc|}
\hline
$L$ & $\eta_s$ & $\phi$ & $\eta_c$ & $J/\psi$ \\
\hline
24 & 0.9705(6) & 0.9471(11) & 1.013(6) & 0.989(5) \\
32 & 0.9737(5) & 0.9536(11) & 1.020(5) & 0.996(6) \\
48 & 0.9774(8) & 0.9597(22) & -- & -- \\
\hline
\end{tabular}
\caption{
The  calculated ``speed of light'' of the $\eta_s$, $\phi$, $\eta_c$ and $J/\psi$  
extracted from each volume using a quadratic fit.
The statistical and systematic uncertainties have been combined in quadrature.
}
\label{tab:c}
\end{table}

As stated previously, the calculations have been performed at only one lattice spacing.
Given that the clover action has been used, lattice-spacing artifacts are expected to be 
 small, scaling as ${\cal O}(a^2,\alpha_s a)$.   
However, the uncertainties in the binding energies introduced by the discretization remain to be quantified, 
and calculations with other ensembles with smaller lattice spacings will be  required
in order to perform a continuum extrapolation.

\section{Nucleus-Quarkonium Binding Energies}
\label{sec:results}

The energies of quarkonium-nucleus systems may be extracted from two-point correlation functions with the appropriate quantum numbers. For the systems of interest, we considered the two-point functions
\begin{eqnarray}
{\cal C}_{{\cal A}{\cal B}}(t)  
&=& 
\left\langle 0 \left| \chi_{\cal A}(t)  \tilde \chi^\dagger_{\cal B}(0)  \right| 0 \right\rangle\,,
\nonumber \\
{\cal C}_{{A}{B}}(t)  
&=& 
\left\langle 0 \left| \chi_{A}(t)  \tilde \chi^\dagger_{B}(0)  \right| 0 \right\rangle\,,
\\
{\cal C}_{\Gamma}(t)  
&=& 
\left\langle 0 \left| \chi_{\overline{Q} \Gamma Q}(t)  \tilde \chi^\dagger_{\overline{Q}\Gamma Q}(0)  \right| 0 \right\rangle\,,
\nonumber 
\end{eqnarray}
with
$\chi_{\cal A} = \chi_A \ \chi_{\overline{Q}\Gamma Q}$
where 
$ \chi_A$ ($\tilde \chi^\dagger_A$) and $\chi_{\overline{Q} \Gamma Q}$ ($\tilde \chi^\dagger_{\overline{Q} \Gamma Q}$) are
 interpolating operators that annihilate (create) states with the quantum numbers of the nucleus $A$ 
 and quarkonia $\overline{Q} \Gamma Q$, respectively (with $\Gamma$ the relevant Dirac structure).\footnote{The calculations presented here ignore the annihilation-type contractions in the quarkonium correlators as they are numerically  expensive to evaluate. These effects are suppressed by the heavy quark mass and are found to be small for charmonium \cite{Levkova:2010ft}. For the strange quarkonium, the effects may be slightly larger and remain to be quantified.} 
 For brevity, the momentum labels on the correlation functions and interpolators are suppressed, 
 however, correlation functions with zero total momentum, as well as those 
 with total momenta $|\frac{L}{2\pi}{\bf P}_{\rm tot}|^2=1,2,3$, are considered.
The  correlation functions can be expanded over the complete set of lattice energy eigenstates with
the appropriate quantum numbers,
\begin{equation}
{\cal C}_{{\cal AB}}(t) =
   \sum_n \ { Z}_{n,{\cal A}} \ { Z}_{n,{\cal B}}^\ast \ e^{-E_n (t_f-t_i)},
 \label{eq:two-pt-correlators}
\end{equation}
where 
the summation is over all eigenstates that couple to the operators $\chi_{\cal A}$, $\chi_{\cal B}$,  
with amplitudes ${ Z}_{n,{\cal A}}$, ${ Z}^\ast_{n,{\cal B}}$.

\begin{figure}[!t]
\includegraphics[width=0.45\textwidth]{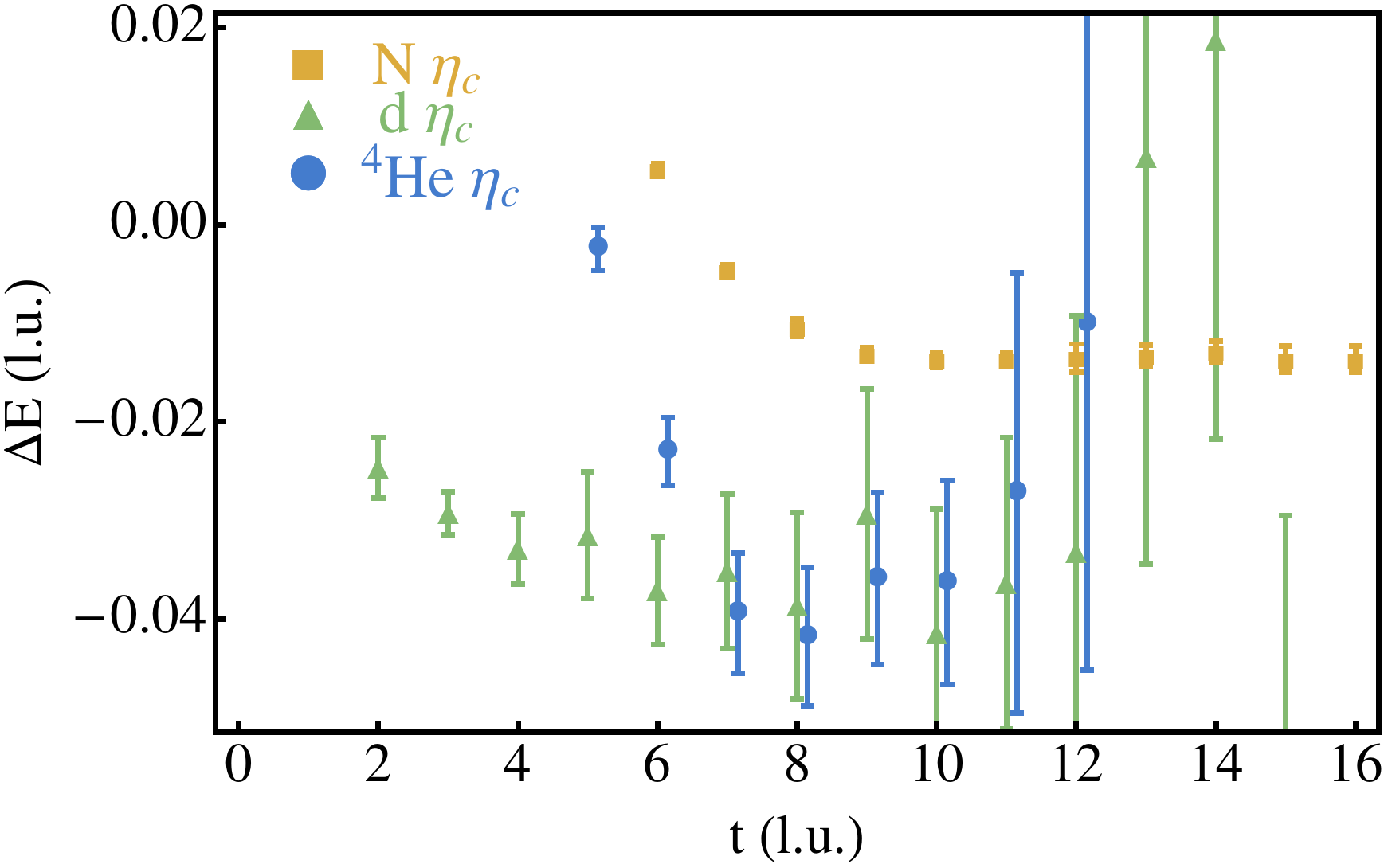}
\caption{
Representative effective energy-shift plots associated with the $N \eta_c$, $d\, \eta_c$ and 
$^4{\rm He}\, \eta_c$ systems obtained from one set of correlation functions  in the $L=32$ ensemble.
}
\label{fig:empcombo}
\end{figure}

In extracting the quarkonium-nucleus
binding energies from the correlation functions,
it is helpful to consider both one-state and two-state fitting functions, truncating the sum in Eq.~(\ref{eq:two-pt-correlators}) to one or two terms. 
At short times, the correlation functions are contaminated by excited states, while at 
later times, the signal-to-noise ratio degrades exponentially. Two-state fits are applicable at earlier times (where the data are more precise) than one-state fits, but the latter serve as an important comparison to understand the systematic uncertainty induced by the choice of the fitting form. 
Performing two-state fits to the single hadron correlation functions yield energy splittings that are consistent with 
the lowest-lying excitation for each species.
In addition to  fits to the two-point correlation functions, 
the binding energy can be isolated by taking ratios of the two-point correlation functions of the system and its components 
(note that in this context, the entire nucleus is considered to be a single component of the system). 
In this latter case, 
the fitting function at large times (neglecting excited states) reduces to
\begin{equation}
{\cal R}(t)=\frac{C_{\cal AB}(t)}{C_{AB}(t)C_{\overline{Q} \Gamma Q}(t)}\rightarrow
   Z e^{-(E_{12}-(E_1+E_2)) (t_f-t_i)},
 \label{eq:two-pt-ratio}
\end{equation}
where $E_{12}$ is the total energy of the ground-state system,
$E_1$ and $E_2$ are the energies of the system components,
and $Z$ is an overall normalization factor. 
The  difference $E_{12}-(E_1+E_2)$ may be fit by a single parameter. 
The statistical quality of the calculations is illustrated in Figure~\ref{fig:empcombo}, 
where  the effective energy-shift plots associated with one of the correlation functions 
for each of the $N \eta_c$, 
$d\, \eta_c$ and 
$^4{\rm He}\, \eta_c$ 
are shown.
These are
derived from sets of correlation functions for which the nucleons are generated from Gaussian-smeared sources and sinks, and the
$\eta_c$ is also derived from a (different) Gaussian-smeared source and sink.
The correlation functions of the quarkonium states have been translated back in time by a small number of time slices,
as was used in Ref.~\cite{Detmold:2008bw}, 
so that the 
start of the plateau regions of the nuclear and quarkonia 
correlation functions
approximately coincide. 
While this does slightly degrade the uncertainty, the fact that the ground-state energies of the quarkonia are more than an order of 
magnitude more precise than those of the nuclei, this time translation has a minimal impact upon the analysis of binding energies.
The fitting intervals used to extract the quarkonium-nucleus 
binding energies from the ratios of correlation functions corresponded approximately  
to those used to extract the binding energies of the nucleus, 
as detailed in Ref.~\cite{Beane:2012vq}.
For the two-state fits, the intervals extend to shorter times by a number of time slices, dependent upon the goodness of fit.
Variations of these fitting intervals are used to estimate the systematic uncertainties associated with extracted fit parameters.

\begin{figure}[!t]
\includegraphics[width=0.45\textwidth]{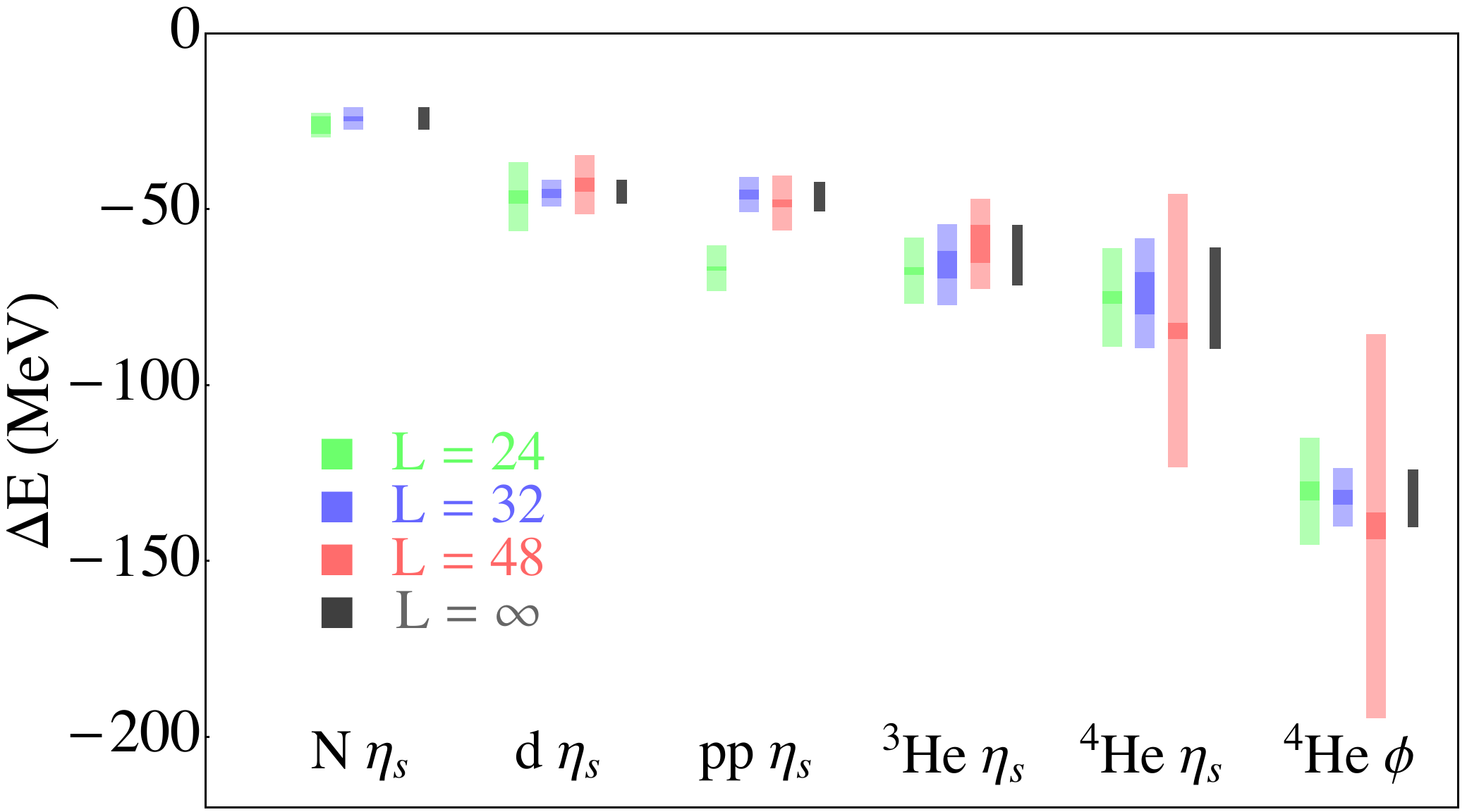}
\includegraphics[width=0.45\textwidth]{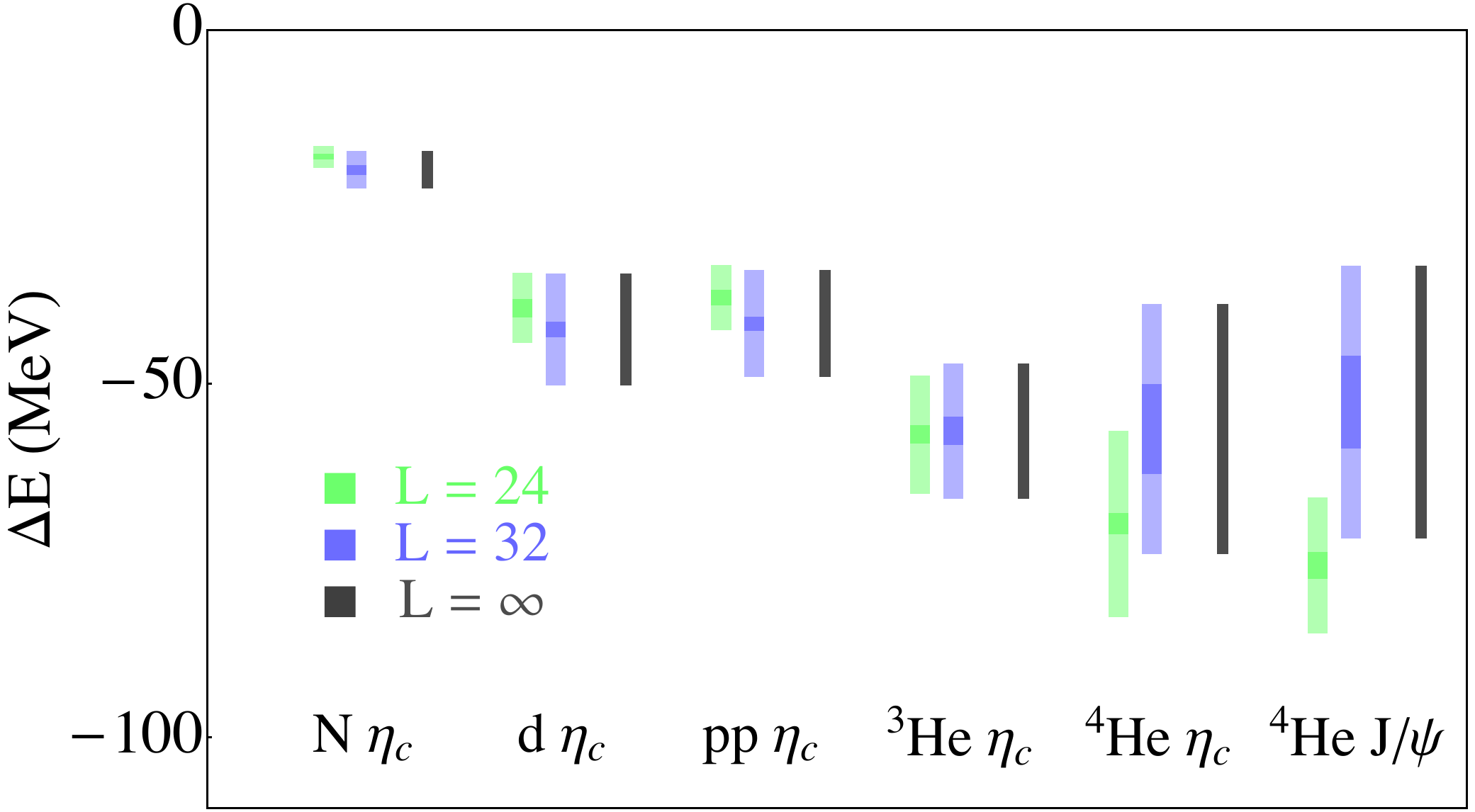}
\caption{
Binding energies of strangeonium-nucleus (upper panel) and charmonium-nucleus (lower panel) systems from Table~\protect\ref{tab:strange-method-vol} and 
Table~\protect\ref{tab:charm-method-vol}.
The inner bands correspond to the statistical uncertainty, while the outer bands correspond to the statistical and systematic uncertainties combined in quadrature.
The right most (gray) band for each system corresponds to the infinite-volume estimate, resulting from a weighted average of the $L=32$ and $L=48$ (where available) 
energies.
}
\label{fig:charm-eff}
\end{figure}

The results of our calculations in the three volumes,  
combining the output from the three analysis methods outlined previously, 
are summarized in  Figure \ref{fig:charm-eff} and in Table~\ref{tab:strange-method-vol} 
for the strangeonium-nucleus systems and in Table~\ref{tab:charm-method-vol} for charmonium-nucleus systems.
The results obtained from one- and two-state fits to the correlation functions 
are consistent with those extracted from fitting to the effective mass at intermediate times, 
but are found to be more precise. 
A systematic fitting uncertainty is assessed based on the differences between the three methods.
\begin{table*}[!ht]
\begin{tabular}{|c|c|c|c|c|}
\hline
System & $24^3 \times 64$ & $32^3 \times 64$ & $48^3 \times 64$ & $L=\infty$ \\
\hline
$N\, \eta_s$ & 26.1(2.5)(2.5) & 24.3(0.7)(3.2) & -  & 24.3(3.2)\\
$d\, \eta_s$  & 46.5(1.9)(9.7) & 45.5(1.3)(3.6) & 43.0(2.0)(8.2) & 45.0(3.5)\\
$pp\, \eta_s$ & 66.9(0.7)(6.5) & 45.8(1.4)(4.8) & 48.3(1.1)(7.7) & 46.5(4.2)\\
$^3{\rm He}\, \eta_s$ & 67.6(1.1)(9.4) & 66(04)(11) & 60(05)(12) & 63.2(8.6)  \\
$^4{\rm He}\, \eta_s$ & 75(02)(14) & 74(06)(14) & 85 (02)(39) & 75(14) \\
$^4{\rm He}\, \phi$  & 130(03)(15) & 132.0(2.1)(8.1) & 140(04)(55) & 132.1 (8.2)\\
\hline
\end{tabular}
\caption{The binding energies (in MeV) of strangeonium-nucleus systems calculated on the $L=24, 32$ and $48$ ensembles.  
The right most column shows the infinite-volume estimate given by the 
weighted average of the $L=32$ and $L=48$ binding energies. 
The first and second set of parentheses show the statistical and quadrature-combined statistical plus systematic uncertainties, respectively.
}
\label{tab:strange-method-vol}
\end{table*}
\begin{table*}[!ht]
\begin{tabular}{|c|c|c|c|}
\hline
System & $24^3 \times 64$ & $32^3 \times 64$ &  $L=\infty$ \\
\hline
$N\, \eta_c$ & 17.9(0.4)(1.5) & 19.8(0.7)(2.6) & 19.8(2.6) \\
$d\, \eta_c$  & 39.3(1.3)(4.8) & 42.4(1.1)(7.9)& 42.4(7.9)  \\
$pp\, \eta_c$ & 37.8(1.1)(4.5) & 41.5(1.0)(7.5) & 41.5(7.6)  \\
$^3{\rm He}\, \eta_c$ & 57.2(1.3)(8.3) & 56.7(2.0)(9.4) & 56.7(9.6)  \\
$^4{\rm He}\, \eta_c$ & 70(02)(13)  & 56(06)(17) & 56(18) \\
$^4{\rm He}\, J/\psi$  & 75.7(1.9)(9.4) & 53(07)(18) & 53(19) \\
\hline
\end{tabular}
\caption{The binding energies (in MeV) of charmonium-nucleus systems 
calculated on the $L=24$ and $ 32$ ensembles. 
The right most column shows the infinite-volume estimate, which, without results on the $L=48$  ensemble, is 
taken to be the binding calculated on the $L=32$ ensemble. 
The first and second set of parentheses shows the  statistical and quadrature-combined statistical plus systematic uncertainties, respectively.
}
\label{tab:charm-method-vol}
\end{table*}

Most of the systems we have explored in this work have negligible
finite volume (FV) effects.  For the isolated nuclear systems, the FV
effects, which depend upon the nuclear binding energies, were
quantified for these ensembles by previous calculations~\cite{Beane:2012vq}, 
from which it was determined that such effects are
negligible in the $L=32$ and $L=48$ ensembles.  The volume effects are also negligible for
the isolated mesons, as is clear by explicit comparison of the
dispersion relations extracted from each ensemble, see
Figure~\ref{fig:disp}.  Finally, the calculated binding energies are
sufficiently deep that the energy gap to the nearest state above the
quarkonium-nucleus ground state is large enough so that the FV
modifications to the binding energy of the combined system are negligible
in the $L=32$ and $L=48$ volumes, as can be seen from  Figure~\ref{fig:charm-eff} 
(the $L=24$ ensemble shows some small volume dependence in a few systems).  
As
a result, the infinite-volume binding energy is taken to be the
weighted average of the binding energy in the $L=32$ and the $L=48$
ensembles (the largest volume is not available for the
charmonium-nucleus systems, but we assume volume effects in this case
are not larger than those in the corresponding strangeonium-nucleus system
and are thus negligible for the $L=32$ results).  The exponential dependence
upon the spatial extent of the lattice for bound systems, along with the measured energy
scales, allow for an estimate of the infinite-volume binding energy
while introducing a systematic uncertainty that is much smaller than
the statistical and fitting systematic uncertainties.
There is one caveat to this discussion of FV effects, that will be discussed in detail in Section~\ref{sec:boosted}.
It is possible, due to the finite time extent of the plateaus, that the states we have identified are contaminated by low-lying 
scattering states at some  level.  
While the uncertainties in the present results preclude 
a stable power-law extrapolation to infinite-volume, 
by making reasonable assumptions about the 
scattering parameters describing 
their interactions, 
our results indicate that such contaminations are small,
providing energy shifts that are smaller than the quoted uncertainties.

In contrast to the charmonium-nucleus systems, the non-interacting
$\eta_s$-nucleus systems are, up to nuclear binding energy
contributions, degenerate with other states, such as $K$-hypernucleus
states in the SU(3) limit.  From the standpoint of SU(3) flavor
symmetry, the charmonia are singlets (charmonia are also deeply bound relative to the $c\overline{c}$ threshold), while the $\eta_s$ is a
combination of a singlet and an octet. In the latter case, this
complicates the classification of the composite systems.  For example,
as the deuteron transforms in a $\overline{\bf 10}$ of SU(3), the
charmonium-deuteron system is also in a $\overline{\bf 10} $
representation, while the strangeonium-deuteron system transforms as
$\left( {\bf 1} \oplus {\bf 8} \right)\otimes \overline{\bf 10} =
 {\bf 8}\oplus 2\cdot\overline{\bf 10}\oplus {\bf 27} \oplus \overline{\bf 35}$.  Including interactions, the energy
eigenvalues of the $\eta_s$-nucleus systems therefore result from
diagonalizing a coupled channels system, and one may anticipate
potential difficulties in extracting the binding energy because of nearby
levels.  {\it A posteriori}, we find that the correlators exhibit
single exponential behavior (to the level at which we can resolve it)
and the corresponding ground states are sufficiently isolated to
permit their extraction.  Physically, the binding of quarkonium to the
nucleus introduces a relatively large energy scale into the
coupled-channel system, leading to an isolated ground state.

\begin{figure}
\includegraphics[width=0.45\textwidth]{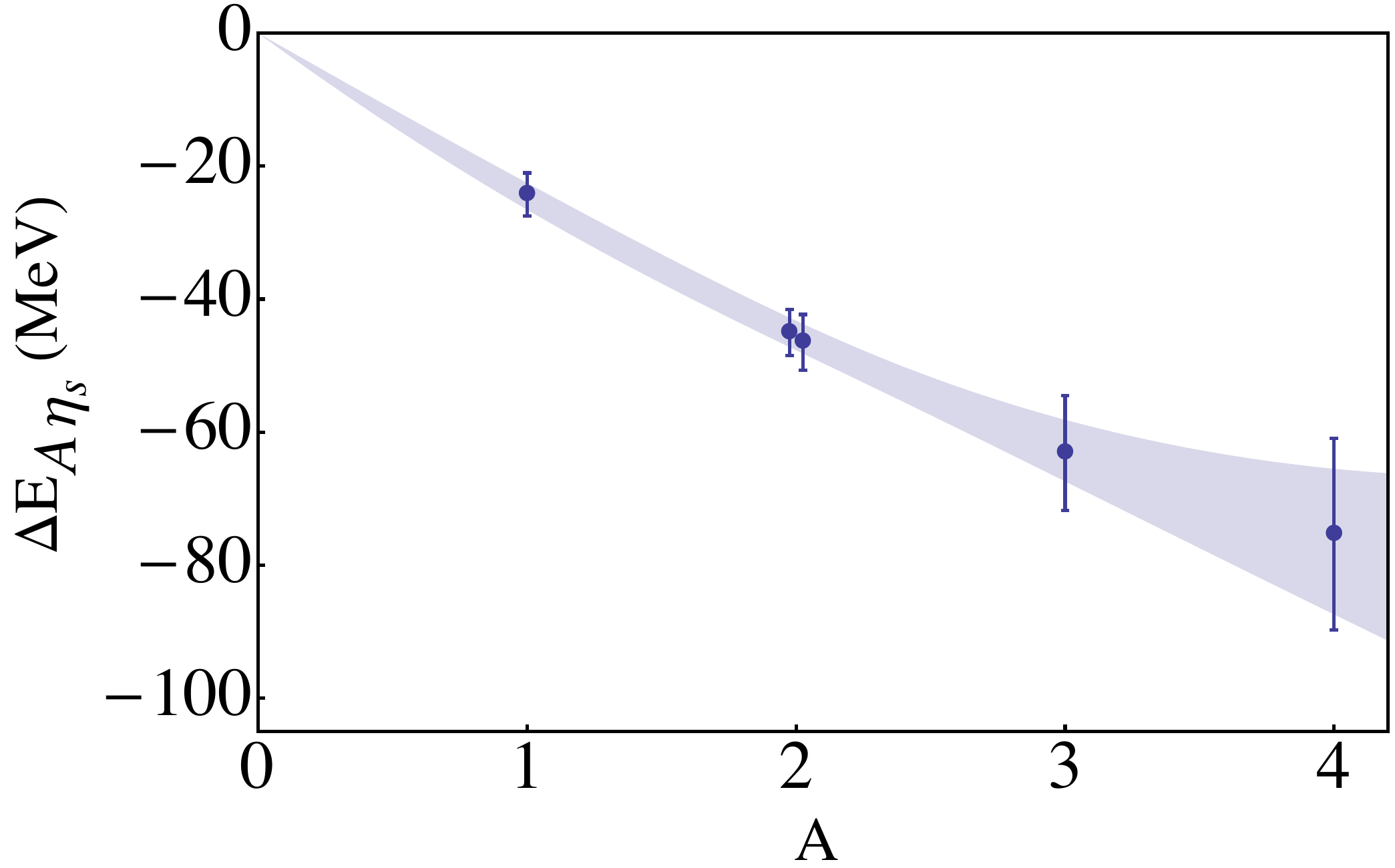}\hspace*{2mm}\; \\
\includegraphics[width=0.44\textwidth]{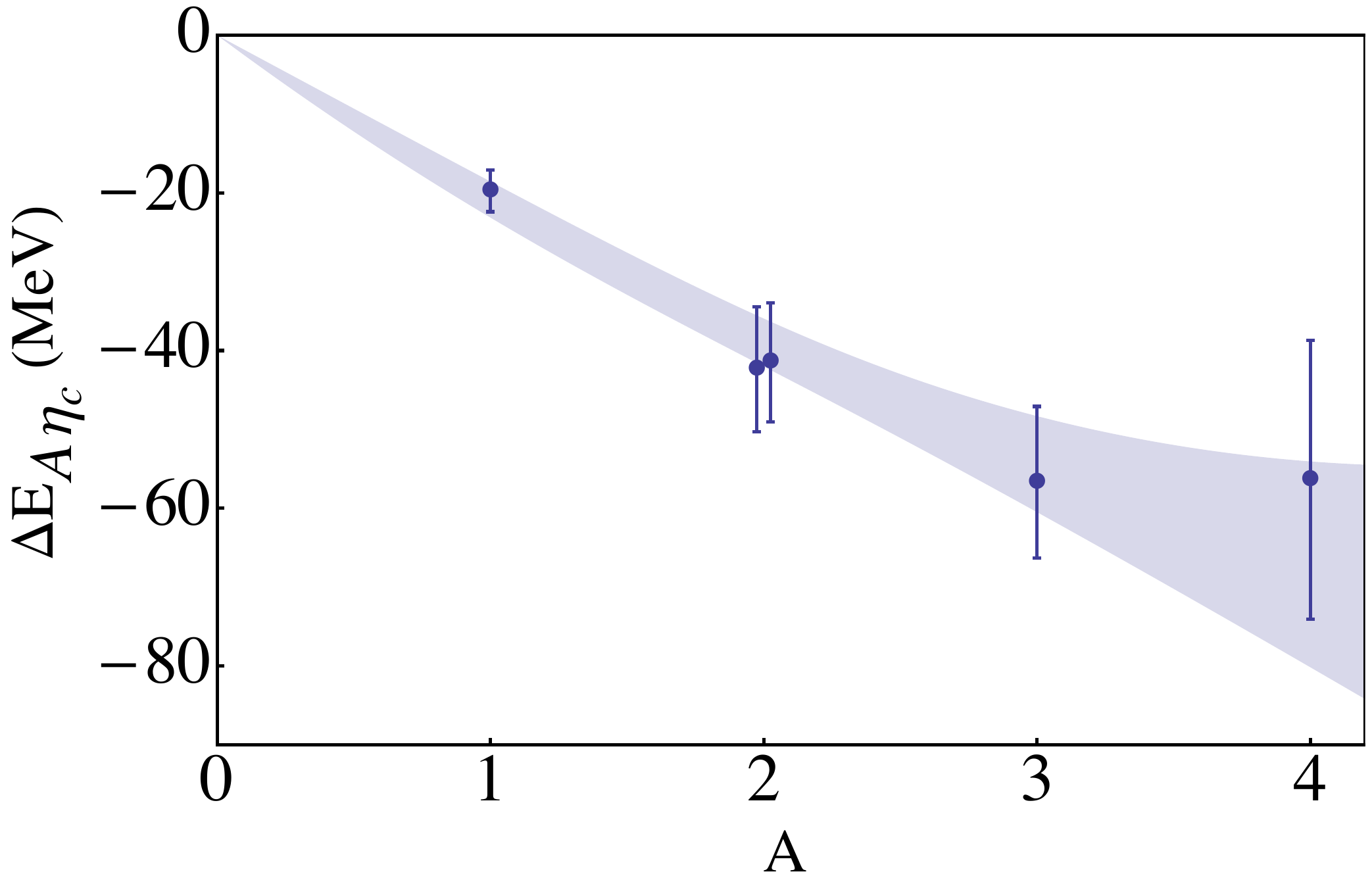}
\caption{
Binding energies of the $A$ $\eta_s$ (upper) and $A$ $\eta_c$ (lower) systems as functions of atomic number. 
For A = 2, we display both the deuteron and $nn$ results.
The shaded region corresponds to a phenomenological  quadratic fit to the results. }
\label{fig:vs-A}
\end{figure}

All of the quarkonium-nucleus systems that we have explored are found
to have binding energies that differ significantly from zero,
and the results are summarized in Figure~\ref{fig:vs-A}.  These binding
energies are quite large when compared to typical nuclear binding
energies at the physical point ($\sim 8~{\rm MeV}$ per nucleon in NM),
but similar in size to the nuclear bindings found at these unphysically
heavy quark masses~\cite{Beane:2012vq}.  In analogy with the liquid-drop model
description of nuclei, where binding energies per nucleon are of the
form $B/A \sim \alpha_V -\alpha_S A^{-1/3}$ (we keep only the volume 
and surface terms with coefficients $\alpha_{V,S}$, respectively), the binding between quarkonia and nuclei
is expected 
to have a similar classical expansion of the form
$B_{A\overline{Q}Q}  \sim \alpha^{\overline{Q}Q}_V  - \alpha^{\overline{Q}Q}_S A^{-1/3}$.
As the long range component of the interaction 
between quarkonia and the nucleons scales as $ V(r) \sim e^{-2 M_\pi r}/r^\alpha$ (with some positive constant $\alpha$), the force is expected to saturate more rapidly with increasing 
nuclear size than for pure nuclear bindings.
Within significant uncertainties, we find the $\eta_c$ to have equal binding to $^3$He and $^4$He, the weighted average of which yields an 
estimate of the nuclear matter binding energy
of $B^{\rm NM} \sim 60~{\rm MeV}$ at this heavy pion mass.
However, we have an insufficient 
range of nuclei to determine if, in fact, the $A=4$ system is at saturation,
so this value is  speculative.

  The leading behavior
of the binding to nuclear matter in the heavy-quark
limit~\cite{Luke:1992tm} is linear in the mass density of the nuclear
system, which itself depends approximately upon the nucleon mass and
baryon number density. Using the experimental nucleon mass, and
assuming the number density is either constant or decreases towards
the physical quark mass, this yields an upper bound on the $\eta_c$
binding energy of $B^{\rm NM}_{\rm phys} \lsim 40~{\rm MeV}$, but without
a full quantification of uncertainties.

\section{Boosted Systems}
\label{sec:boosted}

\begin{figure}
\includegraphics[width=0.45\textwidth]{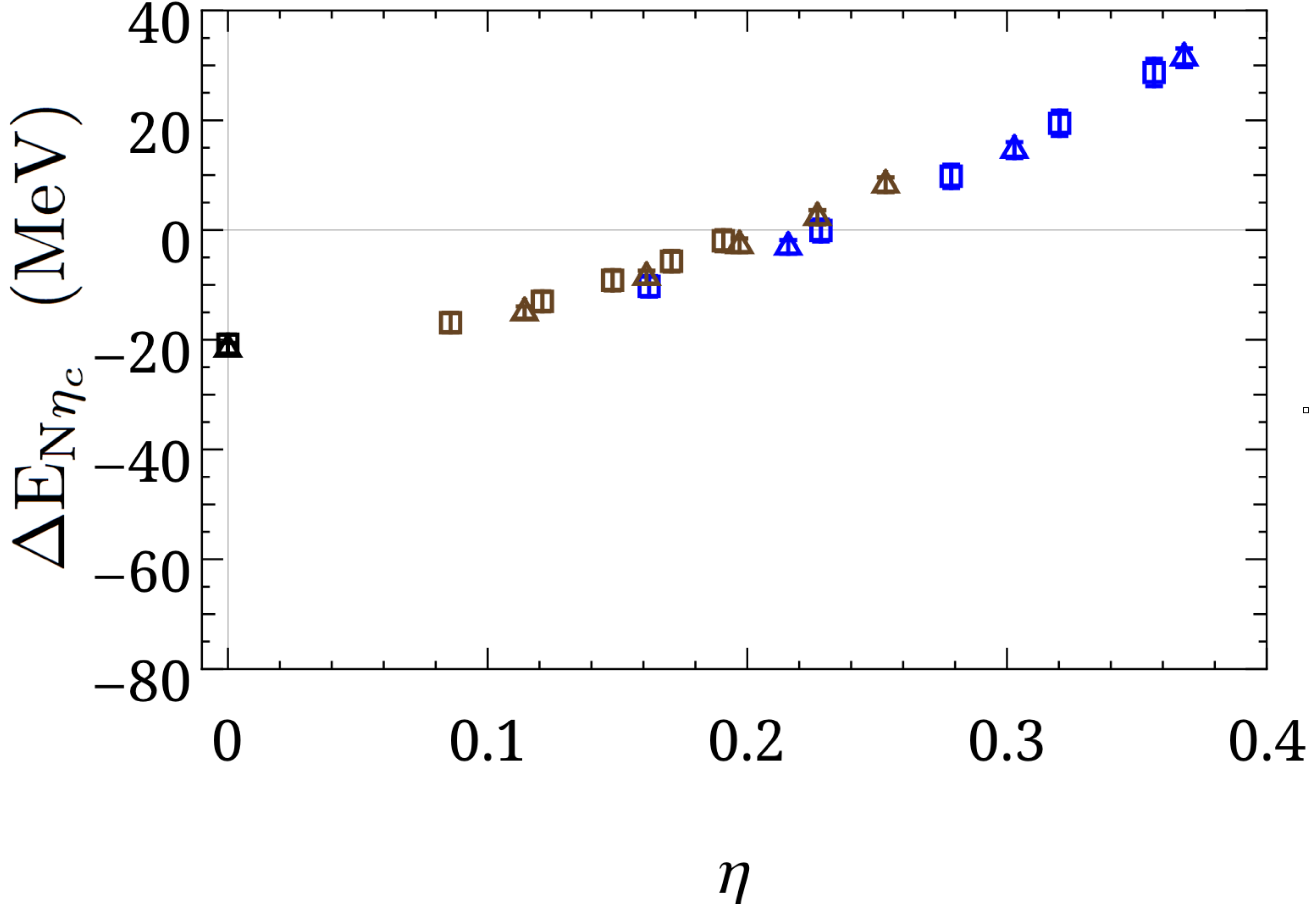} 
\caption{
An example of the energy differences (in MeV) for  charmonium-nucleus systems,
N $\eta_c$,
versus the rapidity of the boosted hadron. 
The brown points show the extracted energies of systems produced from sinks for which the quarkonium is boosted and the nucleon is at rest, 
while the blue points show the extracted energies of systems produced from sinks for which the nucleon is boosted and the quarkonium is at rest The black point correspond to the system produced at rest. 
Triangles (squares) denote results from lattice volumes with spatial extent $L=24$ ($L=32$).
  }
\label{fig:charmboost-eff}
\end{figure}

As discussed previously, quarkonium-nucleus correlation functions associated with a given total
three momentum were constructed by multiplying the appropriate
correlation functions.  In our calculations, at least one of the
component systems was at rest in the lattice volume.  For systems with total
center-of-mass (CoM) momentum, ${\bf P}_{\rm tot}\ne0$, the total
energy of the ground state was translated to the CoM energy, and then
to the binding energy of the system by removing the rest masses of the
constituents.  An example of the energy shifts for the
charmonium-nucleus systems in the CoM frame is shown in
Figure~\ref{fig:charmboost-eff} as a function of relative rapidity,
$\eta = \tanh^{-1}\beta$, where $\beta$ is the velocity of the boosted
hadron.  Similar dependence is seen for all of the quarkonium-nucleus systems that we have studied.
Na\"ively, one expects that the CoM energy should be
independent of the relative velocity, however, this is not what we
find in our results.  Instead, there is a trend for the extracted
total energy to increase approximately quadratically with the relative
rapidity.  We speculate that this behavior arises because the overlap
of the momentum projected sink interpolators  onto a bound state is suppressed at
non-zero relative momentum, while the overlap onto the continuum
states remains of order unity, dictated by the lattice volume.  While
the bound state dominates the 
correlation functions
for $\beta\sim 0$, 
its contribution will be suppressed for interpolating operators
with relative momenta that are of order or greater than the binding
momentum of the state.  At intermediate times from the source, the
effective mass plots associated with such systems may exhibit a
``plateau'' with an energy that exceeds the actual energy of the bound
state.  Toy models of such systems, with two or more nearby states,
can be readily constructed that exhibit such behavior, and there are
sets of natural-sized parameters that are consistent with the behavior
seen in the numerical results.  Only at very large times can the true
ground state be extracted, but at these times the signal-to-noise
ratio has degraded to the point where the energy cannot be usefully
constrained at the current (and foreseeable) statistical precision.
The observed approximate linearity in $\beta^2$ is consistent with
this scenario, but our argument remains a conjecture at this point.
In order to convincingly diagnose the origin of this momentum
dependence, a more extensive set of calculations are required,
involving single- and multi-hadron sources and sinks, and utilizing
the full machinery of the variational method~\cite{Michael:1985ne,
  Luscher:1990ck}.

Our current understanding of the observed relative-velocity dependence
of the extracted binding energies of the quarkonium-nucleus systems
remains incomplete and it is possible that these concerns also effect
the zero velocity systems. 
The associated systematic uncertainties
must be more concretely quantified in future calculations, however the
relatively weak dependence on $\beta$ near $\beta=0$, and the lack of 
volume dependence, suggests that the ground states of these systems
are bound states rather than scattering states.  
From the energies
extracted at non-zero relative velocity, we expect that removing this
systematic will lead to a deeper binding energy than we have
estimated, but within the quoted uncertainties.
To demonstrate the validity of this statement we consider $N$-$\eta_c$ system.
With binding energies in only two volumes, a generic extrapolation of the form
$B(L) = B_0 + \beta/L^3$, 
that would describe such contamination from the lowest-lying continuum state (with an admixture $\beta$),
is unstable when fit to the results, due to the relative size of the uncertainties in each.
However, assuming that the scattering parameters of the system are of natural size, 
and that the extracted energies are perturbatively close to the true binding energy,
the scattering length of this system is found to be 
$a\sim 1~{\rm fm}$ when higher order terms in the effective range expansion are ignored.
This value then yields an expected energy difference between the lowest-lying continuum 
states in the $L=24$ and $32$ volumes of
$\delta E\sim 0.005~{\rm l.u.}\sim 7~{\rm MeV}$ (using L\"uscher's method).
This is larger 
than the difference in ground state energies extracted from the two volumes, 
$\sim 2~{\rm MeV}$, 
indicating that the admixture of scattering state in the observed bound state is small.
Taking  central values to constrain the scattering state contamination, 
the binding energy is $\sim 1.5~{\rm MeV}$ deeper than shown in 
Table~\ref{tab:charm-method-vol}.  
This value is within the uncertainty associated with this binding.
However, the contamination is consistent with zero in all systems we have calculated, 
and this effect should be considered as an uncertainty, smaller than those from other sources,
as opposed to an energy shift.
Further, it can only lead to the extrapolated  binding energies being deeper than shown in 
Table~\ref{tab:charm-method-vol}.  
Only higher precision calculations in additional volumes can further address this issue.

\section{Conclusions}
\label{sec:end}

In this study, we have performed lattice QCD calculations that
demonstrate the existence of bound quarkonium-nucleus systems in QCD
at the flavor-symmetric SU(3) point.  Calculations were performed in
multiple lattice volumes to enable an exploration of volumes effects,
in particular to distinguish between scattering states and bound
states.  Only one lattice spacing was used in this work, and so the
continuum limit could not be taken, however, given the ${\cal O}(a)$
improvement of the lattice action, we expect lattice artifacts to be
smaller than the other uncertainties in our calculation.  
For all of the strangeonium-nucleus and charmonium-nucleus systems 
that we study (atomic numbers $A=1,\ldots,4$), we find significant binding 
 at light quark masses corresponding to 
$M_\pi= M_K\sim 805$ MeV. Assuming the consistency of the bindings 
for $A=3$ and 4 is indicative of saturation of the interactions, we 
infer a charmonium-nuclear matter binding energy
of $B^{\rm NM} \sim 60~{\rm MeV}$ at this heavy pion mass, although 
further studies are required to confirm saturation.

As the quark masses decrease towards their physical values, 
the nucleon mass decreases and it is also expected that the energy 
density of a nucleus will decrease \cite{Beane:2012vq}. Quarkonium-nucleus systems are 
therefore likely to  be less bound at lighter quark 
masses and it is possible that 
the systems involving the lightest nuclei will be unbound at the
physical point.  Additional lattice QCD calculations at smaller
light-quark masses will be necessary to investigate whether this is
the case.  The clean signals found in this study at the SU(3) point,
suggest that such studies will be able to conclusively resolve the
nature of a range of quarkonium-nucleus systems.
For the case of nuclear matter, assuming our numerical results for the charmonium-nucleus binding energies indicate saturation, the leading order extrapolation to the physical quark masses results in an 
estimated binding energy of $B^{\rm NM}_{\rm phys} \lsim 40~{\rm MeV}$, although the uncertainties
in this result are not yet fully quantified.
 With greater computational
resources becoming available, future calculations will be more
precise, extended to larger nuclei, and be will performed at smaller
lattice spacings, which will ultimately lead to predictions for the
binding of quarkonium to nuclei that can guide, and be directly
compared with, ongoing and future experiments.


\

\begin{acknowledgments}
  We would like to thank Zohreh Davoudi for many interesting
  discussions relating to this project, and Tom Luu and Andre
  Walker-Loud for collaboration in related work.  Calculations were
  carried using computational resources provided by the Extreme
  Science and Engineering Discovery Environment (XSEDE), which is
  supported by National Science Foundation grant number OCI-1053575,
  and NERSC (supported by U.S. Department of Energy Grant Number
  DE-AC02-05CH11231), and by the USQCD collaboration. Additional
  calculations were performed at the HYAK facility at the University
  of Washington.  The calculations used the Chroma software
  suite~\cite{Edwards:2004sx}.  SRB was partially supported by NSF
  continuing grant PHY1206498.  WD was supported by the
  U.S. Department of Energy Early Career Research Award DE-SC0010495
  and the Solomon Buchsbaum Fund at MIT.  HWL was supported in part by
  DOE grant No.~DE-FG02-97ER4014.  KO was supported by the
  U.S. Department of Energy through Grant Number DE- FG02-04ER41302
  and through Grant Number DE-AC05-06OR23177 under which JSA operates
  the Thomas Jefferson National Accelerator Facility.  The work of AP
  was supported by the contract FIS2011-24154 from MEC (Spain) and
  FEDER. MJS was supported in part by DOE grant No.~DE-FG02-00ER41132.
\end{acknowledgments}

\bibliography{test}

\begin{thebibliography}{41}
\expandafter\ifx\csname natexlab\endcsname\relax\def\natexlab#1{#1}\fi
\expandafter\ifx\csname bibnamefont\endcsname\relax
  \def\bibnamefont#1{#1}\fi
\expandafter\ifx\csname bibfnamefont\endcsname\relax
  \def\bibfnamefont#1{#1}\fi
\expandafter\ifx\csname citenamefont\endcsname\relax
  \def\citenamefont#1{#1}\fi
\expandafter\ifx\csname url\endcsname\relax
  \def\url#1{\texttt{#1}}\fi
\expandafter\ifx\csname urlprefix\endcsname\relax\def\urlprefix{URL }\fi
\providecommand{\bibinfo}[2]{#2}
\providecommand{\eprint}[2][]{\url{#2}}

\bibitem[{\citenamefont{Brodsky et~al.}(1990)\citenamefont{Brodsky, Schmidt,
  and de~Teramond}}]{Brodsky:1989jd}
\bibinfo{author}{\bibfnamefont{S.~J.} \bibnamefont{Brodsky}},
  \bibinfo{author}{\bibfnamefont{I.}~\bibnamefont{Schmidt}}, \bibnamefont{and}
  \bibinfo{author}{\bibfnamefont{G.}~\bibnamefont{de~Teramond}},
  \bibinfo{journal}{Phys.Rev.Lett.} \textbf{\bibinfo{volume}{64}},
  \bibinfo{pages}{1011} (\bibinfo{year}{1990}).

\bibitem[{\citenamefont{Wasson}(1991)}]{Wasson:1991fb}
\bibinfo{author}{\bibfnamefont{D.}~\bibnamefont{Wasson}},
  \bibinfo{journal}{Phys.Rev.Lett.} \textbf{\bibinfo{volume}{67}},
  \bibinfo{pages}{2237} (\bibinfo{year}{1991}).

\bibitem[{\citenamefont{Luke et~al.}(1992)\citenamefont{Luke, Manohar, and
  Savage}}]{Luke:1992tm}
\bibinfo{author}{\bibfnamefont{M.~E.} \bibnamefont{Luke}},
  \bibinfo{author}{\bibfnamefont{A.~V.} \bibnamefont{Manohar}},
  \bibnamefont{and} \bibinfo{author}{\bibfnamefont{M.~J.}
  \bibnamefont{Savage}}, \bibinfo{journal}{Phys.Lett.}
  \textbf{\bibinfo{volume}{B288}}, \bibinfo{pages}{355} (\bibinfo{year}{1992}),
  \eprint{hep-ph/9204219}.

\bibitem[{\citenamefont{Brodsky and Miller}(1997)}]{Brodsky:1997gh}
\bibinfo{author}{\bibfnamefont{S.~J.} \bibnamefont{Brodsky}} \bibnamefont{and}
  \bibinfo{author}{\bibfnamefont{G.~A.} \bibnamefont{Miller}},
  \bibinfo{journal}{Phys.Lett.} \textbf{\bibinfo{volume}{B412}},
  \bibinfo{pages}{125} (\bibinfo{year}{1997}), \eprint{hep-ph/9707382}.

\bibitem[{\citenamefont{de~Teramond et~al.}(1998)\citenamefont{de~Teramond,
  Espinoza, and Ortega-Rodriguez}}]{deTeramond:1997ny}
\bibinfo{author}{\bibfnamefont{G.~F.} \bibnamefont{de~Teramond}},
  \bibinfo{author}{\bibfnamefont{R.}~\bibnamefont{Espinoza}}, \bibnamefont{and}
  \bibinfo{author}{\bibfnamefont{M.}~\bibnamefont{Ortega-Rodriguez}},
  \bibinfo{journal}{Phys.Rev.} \textbf{\bibinfo{volume}{D58}},
  \bibinfo{pages}{034012} (\bibinfo{year}{1998}), \eprint{hep-ph/9708202}.

\bibitem[{\citenamefont{Lee and Ko}(2003)}]{Ko:2000jx}
\bibinfo{author}{\bibfnamefont{S.~H.} \bibnamefont{Lee}} \bibnamefont{and}
  \bibinfo{author}{\bibfnamefont{C.}~\bibnamefont{Ko}},
  \bibinfo{journal}{Phys.Rev.} \textbf{\bibinfo{volume}{C67}},
  \bibinfo{pages}{038202} (\bibinfo{year}{2003}), \eprint{nucl-th/0208003}.

\bibitem[{\citenamefont{Tsushima et~al.}(2011)\citenamefont{Tsushima, Lu,
  Krein, and Thomas}}]{Tsushima:2011kh}
\bibinfo{author}{\bibfnamefont{K.}~\bibnamefont{Tsushima}},
  \bibinfo{author}{\bibfnamefont{D.}~\bibnamefont{Lu}},
  \bibinfo{author}{\bibfnamefont{G.}~\bibnamefont{Krein}}, \bibnamefont{and}
  \bibinfo{author}{\bibfnamefont{A.}~\bibnamefont{Thomas}},
  \bibinfo{journal}{Phys.Rev.} \textbf{\bibinfo{volume}{C83}},
  \bibinfo{pages}{065208} (\bibinfo{year}{2011}), \eprint{1103.5516}.

\bibitem[{\citenamefont{Yokota et~al.}(2013)\citenamefont{Yokota, Hiyama, and
  Oka}}]{Yokota:2013sfa}
\bibinfo{author}{\bibfnamefont{A.}~\bibnamefont{Yokota}},
  \bibinfo{author}{\bibfnamefont{E.}~\bibnamefont{Hiyama}}, \bibnamefont{and}
  \bibinfo{author}{\bibfnamefont{M.}~\bibnamefont{Oka}},
  \bibinfo{journal}{PTEP} \textbf{\bibinfo{volume}{2013}},
  \bibinfo{pages}{113D01} (\bibinfo{year}{2013}), \eprint{1308.6102}.

\bibitem[{\citenamefont{Yamagata-Sekihara
  et~al.}(2010)\citenamefont{Yamagata-Sekihara, Cabrera, Vicente~Vacas, and
  Hirenzaki}}]{YamagataSekihara:2010rb}
\bibinfo{author}{\bibfnamefont{J.}~\bibnamefont{Yamagata-Sekihara}},
  \bibinfo{author}{\bibfnamefont{D.}~\bibnamefont{Cabrera}},
  \bibinfo{author}{\bibfnamefont{M.~J.} \bibnamefont{Vicente~Vacas}},
  \bibnamefont{and}
  \bibinfo{author}{\bibfnamefont{S.}~\bibnamefont{Hirenzaki}},
  \bibinfo{journal}{Prog.Theor.Phys.} \textbf{\bibinfo{volume}{124}},
  \bibinfo{pages}{147} (\bibinfo{year}{2010}), \eprint{1001.2235}.

\bibitem[{\citenamefont{Haider and Liu}(1986)}]{Haider:1986sa}
\bibinfo{author}{\bibfnamefont{Q.}~\bibnamefont{Haider}} \bibnamefont{and}
  \bibinfo{author}{\bibfnamefont{L.}~\bibnamefont{Liu}},
  \bibinfo{journal}{Phys.Lett.} \textbf{\bibinfo{volume}{B172}},
  \bibinfo{pages}{257} (\bibinfo{year}{1986}).

\bibitem[{ATH()}]{ATHENNA}
\urlprefix\url{http://www.jlab.org/exp\_prog/proposals/12/PR12-12-006.pdf}.

\bibitem[{fai()}]{fair}
\urlprefix\url{http://www.gsi.de/en/research/fair.htm}.

\bibitem[{\citenamefont{Barabanov}(2012)}]{Barabanov:2013cna}
\bibinfo{author}{\bibfnamefont{M.~Y.} \bibnamefont{Barabanov}},
  \bibinfo{journal}{PoS} \textbf{\bibinfo{volume}{Baldin-ISHEPP-XXI}},
  \bibinfo{pages}{111} (\bibinfo{year}{2012}).

\bibitem[{jpa()}]{jparc}
\urlprefix\url{http://j-parc.jp/index-e.html}.

\bibitem[{\citenamefont{Pfeiffer et~al.}(2004)\citenamefont{Pfeiffer, Ahrens,
  Annand, Beck, Caselotti et~al.}}]{Pfeiffer:2003zd}
\bibinfo{author}{\bibfnamefont{M.}~\bibnamefont{Pfeiffer}},
  \bibinfo{author}{\bibfnamefont{J.}~\bibnamefont{Ahrens}},
  \bibinfo{author}{\bibfnamefont{J.}~\bibnamefont{Annand}},
  \bibinfo{author}{\bibfnamefont{R.}~\bibnamefont{Beck}},
  \bibinfo{author}{\bibfnamefont{G.}~\bibnamefont{Caselotti}},
  \bibnamefont{et~al.}, \bibinfo{journal}{Phys.Rev.Lett.}
  \textbf{\bibinfo{volume}{92}}, \bibinfo{pages}{252001}
  (\bibinfo{year}{2004}), \eprint{nucl-ex/0312011}.

\bibitem[{\citenamefont{Budzanowski et~al.}(2009)}]{Budzanowski:2008fr}
\bibinfo{author}{\bibfnamefont{A.}~\bibnamefont{Budzanowski}}
  \bibnamefont{et~al.} (\bibinfo{collaboration}{COSY-GEM Collaboration}),
  \bibinfo{journal}{Phys.Rev.} \textbf{\bibinfo{volume}{C79}},
  \bibinfo{pages}{012201} (\bibinfo{year}{2009}), \eprint{0812.4187}.

\bibitem[{\citenamefont{Beane et~al.}(2011)}]{Beane:2010hg}
\bibinfo{author}{\bibfnamefont{S.}~\bibnamefont{Beane}} \bibnamefont{et~al.}
  (\bibinfo{collaboration}{NPLQCD Collaboration}),
  \bibinfo{journal}{Phys.Rev.Lett.} \textbf{\bibinfo{volume}{106}},
  \bibinfo{pages}{162001} (\bibinfo{year}{2011}), \eprint{1012.3812}.

\bibitem[{\citenamefont{Beane et~al.}(2012)}]{Beane:2011iw}
\bibinfo{author}{\bibfnamefont{S.}~\bibnamefont{Beane}} \bibnamefont{et~al.}
  (\bibinfo{collaboration}{NPLQCD Collaboration}), \bibinfo{journal}{Phys.Rev.}
  \textbf{\bibinfo{volume}{D85}}, \bibinfo{pages}{054511}
  (\bibinfo{year}{2012}), \eprint{1109.2889}.

\bibitem[{\citenamefont{Yamazaki et~al.}(2011)\citenamefont{Yamazaki,
  Kuramashi, and Ukawa}}]{Yamazaki:2011nd}
\bibinfo{author}{\bibfnamefont{T.}~\bibnamefont{Yamazaki}},
  \bibinfo{author}{\bibfnamefont{Y.}~\bibnamefont{Kuramashi}},
  \bibnamefont{and} \bibinfo{author}{\bibfnamefont{A.}~\bibnamefont{Ukawa}}
  (\bibinfo{collaboration}{Collaboration for the PACS-CS}),
  \bibinfo{journal}{Phys.Rev.} \textbf{\bibinfo{volume}{D84}},
  \bibinfo{pages}{054506} (\bibinfo{year}{2011}), \eprint{1105.1418}.

\bibitem[{\citenamefont{Yamazaki et~al.}(2012)\citenamefont{Yamazaki, Ishikawa,
  Kuramashi, and Ukawa}}]{Yamazaki:2012hi}
\bibinfo{author}{\bibfnamefont{T.}~\bibnamefont{Yamazaki}},
  \bibinfo{author}{\bibfnamefont{K.-i.} \bibnamefont{Ishikawa}},
  \bibinfo{author}{\bibfnamefont{Y.}~\bibnamefont{Kuramashi}},
  \bibnamefont{and} \bibinfo{author}{\bibfnamefont{A.}~\bibnamefont{Ukawa}},
  \bibinfo{journal}{Phys.Rev.} \textbf{\bibinfo{volume}{D86}},
  \bibinfo{pages}{074514} (\bibinfo{year}{2012}), \eprint{1207.4277}.

\bibitem[{\citenamefont{Beane et~al.}(2013{\natexlab{a}})\citenamefont{Beane,
  Chang, Cohen, Detmold, Lin et~al.}}]{Beane:2012vq}
\bibinfo{author}{\bibfnamefont{S.}~\bibnamefont{Beane}},
  \bibinfo{author}{\bibfnamefont{E.}~\bibnamefont{Chang}},
  \bibinfo{author}{\bibfnamefont{S.}~\bibnamefont{Cohen}},
  \bibinfo{author}{\bibfnamefont{W.}~\bibnamefont{Detmold}},
  \bibinfo{author}{\bibfnamefont{H.}~\bibnamefont{Lin}}, \bibnamefont{et~al.},
  \bibinfo{journal}{Phys.Rev.} \textbf{\bibinfo{volume}{D87}},
  \bibinfo{pages}{034506} (\bibinfo{year}{2013}{\natexlab{a}}),
  \eprint{1206.5219}.

\bibitem[{\citenamefont{Beane et~al.}(2013{\natexlab{b}})}]{Beane:2013br}
\bibinfo{author}{\bibfnamefont{S.}~\bibnamefont{Beane}} \bibnamefont{et~al.}
  (\bibinfo{collaboration}{NPLQCD Collaboration}), \bibinfo{journal}{Phys.Rev.}
  \textbf{\bibinfo{volume}{C88}}, \bibinfo{pages}{024003}
  (\bibinfo{year}{2013}{\natexlab{b}}), \eprint{1301.5790}.

\bibitem[{\citenamefont{Beane et~al.}(2014)\citenamefont{Beane, Chang, Cohen,
  Detmold, Lin, Orginos, Parreno, Savage, and Tiburzi}}]{Beane:2014ora}
\bibinfo{author}{\bibfnamefont{S.~R.} \bibnamefont{Beane}},
  \bibinfo{author}{\bibfnamefont{E.}~\bibnamefont{Chang}},
  \bibinfo{author}{\bibfnamefont{S.}~\bibnamefont{Cohen}},
  \bibinfo{author}{\bibfnamefont{W.}~\bibnamefont{Detmold}},
  \bibinfo{author}{\bibfnamefont{H.~W.} \bibnamefont{Lin}},
  \bibinfo{author}{\bibfnamefont{K.}~\bibnamefont{Orginos}},
  \bibinfo{author}{\bibfnamefont{A.}~\bibnamefont{Parreno}},
  \bibinfo{author}{\bibfnamefont{M.~J.} \bibnamefont{Savage}},
  \bibnamefont{and} \bibinfo{author}{\bibfnamefont{B.~C.}
  \bibnamefont{Tiburzi}}, \bibinfo{journal}{Phys. Rev. Lett.}
  \textbf{\bibinfo{volume}{113}}, \bibinfo{pages}{252001}
  (\bibinfo{year}{2014}), \eprint{1409.3556}.

\bibitem[{\citenamefont{Detmold and Savage}(2009)}]{Detmold:2008bw}
\bibinfo{author}{\bibfnamefont{W.}~\bibnamefont{Detmold}} \bibnamefont{and}
  \bibinfo{author}{\bibfnamefont{M.~J.} \bibnamefont{Savage}},
  \bibinfo{journal}{Phys.Rev.Lett.} \textbf{\bibinfo{volume}{102}},
  \bibinfo{pages}{032004} (\bibinfo{year}{2009}), \eprint{0809.0892}.

\bibitem[{\citenamefont{Detmold et~al.}(2013)\citenamefont{Detmold, Meinel, and
  Shi}}]{Detmold:2012pi}
\bibinfo{author}{\bibfnamefont{W.}~\bibnamefont{Detmold}},
  \bibinfo{author}{\bibfnamefont{S.}~\bibnamefont{Meinel}}, \bibnamefont{and}
  \bibinfo{author}{\bibfnamefont{Z.}~\bibnamefont{Shi}},
  \bibinfo{journal}{Phys.Rev.} \textbf{\bibinfo{volume}{D87}},
  \bibinfo{pages}{094504} (\bibinfo{year}{2013}), \eprint{1211.3156}.

\bibitem[{\citenamefont{Yokokawa et~al.}(2006)\citenamefont{Yokokawa, Sasaki,
  Hatsuda, and Hayashigaki}}]{Yokokawa:2006td}
\bibinfo{author}{\bibfnamefont{K.}~\bibnamefont{Yokokawa}},
  \bibinfo{author}{\bibfnamefont{S.}~\bibnamefont{Sasaki}},
  \bibinfo{author}{\bibfnamefont{T.}~\bibnamefont{Hatsuda}}, \bibnamefont{and}
  \bibinfo{author}{\bibfnamefont{A.}~\bibnamefont{Hayashigaki}},
  \bibinfo{journal}{Phys.Rev.} \textbf{\bibinfo{volume}{D74}},
  \bibinfo{pages}{034504} (\bibinfo{year}{2006}), \eprint{hep-lat/0605009}.

\bibitem[{\citenamefont{Liu et~al.}(2008)\citenamefont{Liu, Lin, and
  Orginos}}]{Liu:2008rza}
\bibinfo{author}{\bibfnamefont{L.}~\bibnamefont{Liu}},
  \bibinfo{author}{\bibfnamefont{H.-W.} \bibnamefont{Lin}}, \bibnamefont{and}
  \bibinfo{author}{\bibfnamefont{K.}~\bibnamefont{Orginos}},
  \bibinfo{journal}{PoS} \textbf{\bibinfo{volume}{LATTICE2008}},
  \bibinfo{pages}{112} (\bibinfo{year}{2008}), \eprint{0810.5412}.

\bibitem[{\citenamefont{Kawanai and
  Sasaki}(2010{\natexlab{a}})}]{Kawanai:2010cq}
\bibinfo{author}{\bibfnamefont{T.}~\bibnamefont{Kawanai}} \bibnamefont{and}
  \bibinfo{author}{\bibfnamefont{S.}~\bibnamefont{Sasaki}},
  \bibinfo{journal}{AIP Conf.Proc.} \textbf{\bibinfo{volume}{1296}},
  \bibinfo{pages}{294} (\bibinfo{year}{2010}{\natexlab{a}}),
  \eprint{1007.1515}.

\bibitem[{\citenamefont{Kawanai and
  Sasaki}(2010{\natexlab{b}})}]{Kawanai:2010ru}
\bibinfo{author}{\bibfnamefont{T.}~\bibnamefont{Kawanai}} \bibnamefont{and}
  \bibinfo{author}{\bibfnamefont{S.}~\bibnamefont{Sasaki}},
  \bibinfo{journal}{PoS} \textbf{\bibinfo{volume}{LATTICE2010}},
  \bibinfo{pages}{156} (\bibinfo{year}{2010}{\natexlab{b}}),
  \eprint{1011.1322}.

\bibitem[{\citenamefont{Kawanai and Sasaki}(2012)}]{Kawanai:2011jt}
\bibinfo{author}{\bibfnamefont{T.}~\bibnamefont{Kawanai}} \bibnamefont{and}
  \bibinfo{author}{\bibfnamefont{S.}~\bibnamefont{Sasaki}},
  \bibinfo{journal}{Phys.Rev.} \textbf{\bibinfo{volume}{D85}},
  \bibinfo{pages}{091503} (\bibinfo{year}{2012}), \eprint{1110.0888}.

\bibitem[{\citenamefont{L{\"u}scher and Weisz}(1985)}]{Luscher:1984xn}
\bibinfo{author}{\bibfnamefont{M.}~\bibnamefont{L{\"u}scher}} \bibnamefont{and}
  \bibinfo{author}{\bibfnamefont{P.}~\bibnamefont{Weisz}},
  \bibinfo{journal}{Commun.Math.Phys.} \textbf{\bibinfo{volume}{97}},
  \bibinfo{pages}{59} (\bibinfo{year}{1985}).

\bibitem[{\citenamefont{Sheikholeslami and
  Wohlert}(1985)}]{Sheikholeslami:1985ij}
\bibinfo{author}{\bibfnamefont{B.}~\bibnamefont{Sheikholeslami}}
  \bibnamefont{and} \bibinfo{author}{\bibfnamefont{R.}~\bibnamefont{Wohlert}},
  \bibinfo{journal}{Nucl.Phys.} \textbf{\bibinfo{volume}{B259}},
  \bibinfo{pages}{572} (\bibinfo{year}{1985}).

\bibitem[{\citenamefont{Morningstar and Peardon}(2004)}]{Morningstar:2003gk}
\bibinfo{author}{\bibfnamefont{C.}~\bibnamefont{Morningstar}} \bibnamefont{and}
  \bibinfo{author}{\bibfnamefont{M.~J.} \bibnamefont{Peardon}},
  \bibinfo{journal}{Phys.Rev.} \textbf{\bibinfo{volume}{D69}},
  \bibinfo{pages}{054501} (\bibinfo{year}{2004}), \eprint{hep-lat/0311018}.

\bibitem[{\citenamefont{Edwards}()}]{EdwardsPC}
\bibinfo{author}{\bibfnamefont{R.~G.} \bibnamefont{Edwards}},
  \bibinfo{note}{{\it private communication}}.

\bibitem[{\citenamefont{Detmold and Orginos}(2013)}]{Detmold:2012eu}
\bibinfo{author}{\bibfnamefont{W.}~\bibnamefont{Detmold}} \bibnamefont{and}
  \bibinfo{author}{\bibfnamefont{K.}~\bibnamefont{Orginos}},
  \bibinfo{journal}{Phys.Rev.} \textbf{\bibinfo{volume}{D87}},
  \bibinfo{pages}{114512} (\bibinfo{year}{2013}), \eprint{1207.1452}.

\bibitem[{\citenamefont{El-Khadra et~al.}(1997)\citenamefont{El-Khadra,
  Kronfeld, and Mackenzie}}]{ElKhadra:1996mp}
\bibinfo{author}{\bibfnamefont{A.~X.} \bibnamefont{El-Khadra}},
  \bibinfo{author}{\bibfnamefont{A.~S.} \bibnamefont{Kronfeld}},
  \bibnamefont{and} \bibinfo{author}{\bibfnamefont{P.~B.}
  \bibnamefont{Mackenzie}}, \bibinfo{journal}{Phys.Rev.}
  \textbf{\bibinfo{volume}{D55}}, \bibinfo{pages}{3933} (\bibinfo{year}{1997}),
  \eprint{hep-lat/9604004}.

\bibitem[{\citenamefont{Brown et~al.}(2014)\citenamefont{Brown, Detmold,
  Meinel, and Orginos}}]{Brown:2014ena}
\bibinfo{author}{\bibfnamefont{Z.~S.} \bibnamefont{Brown}},
  \bibinfo{author}{\bibfnamefont{W.}~\bibnamefont{Detmold}},
  \bibinfo{author}{\bibfnamefont{S.}~\bibnamefont{Meinel}}, \bibnamefont{and}
  \bibinfo{author}{\bibfnamefont{K.}~\bibnamefont{Orginos}},
  \bibinfo{journal}{Phys.Rev.} \textbf{\bibinfo{volume}{D90}},
  \bibinfo{pages}{094507} (\bibinfo{year}{2014}), \eprint{1409.0497}.

\bibitem[{\citenamefont{Levkova and DeTar}(2011)}]{Levkova:2010ft}
\bibinfo{author}{\bibfnamefont{L.}~\bibnamefont{Levkova}} \bibnamefont{and}
  \bibinfo{author}{\bibfnamefont{C.}~\bibnamefont{DeTar}},
  \bibinfo{journal}{Phys.Rev.} \textbf{\bibinfo{volume}{D83}},
  \bibinfo{pages}{074504} (\bibinfo{year}{2011}), \eprint{1012.1837}.

\bibitem[{\citenamefont{Michael}(1985)}]{Michael:1985ne}
\bibinfo{author}{\bibfnamefont{C.}~\bibnamefont{Michael}},
  \bibinfo{journal}{Nucl.Phys.} \textbf{\bibinfo{volume}{B259}},
  \bibinfo{pages}{58} (\bibinfo{year}{1985}).

\bibitem[{\citenamefont{L{\"u}scher and Wolff}(1990)}]{Luscher:1990ck}
\bibinfo{author}{\bibfnamefont{M.}~\bibnamefont{L{\"u}scher}} \bibnamefont{and}
  \bibinfo{author}{\bibfnamefont{U.}~\bibnamefont{Wolff}},
  \bibinfo{journal}{Nucl.Phys.} \textbf{\bibinfo{volume}{B339}},
  \bibinfo{pages}{222} (\bibinfo{year}{1990}).

\bibitem[{\citenamefont{Edwards and Joo}(2005)}]{Edwards:2004sx}
\bibinfo{author}{\bibfnamefont{R.~G.} \bibnamefont{Edwards}} \bibnamefont{and}
  \bibinfo{author}{\bibfnamefont{B.}~\bibnamefont{Joo}}
  (\bibinfo{collaboration}{SciDAC Collaboration, LHPC Collaboration, UKQCD
  Collaboration}), \bibinfo{journal}{Nucl.Phys.Proc.Suppl.}
  \textbf{\bibinfo{volume}{140}}, \bibinfo{pages}{832} (\bibinfo{year}{2005}),
  \eprint{hep-lat/0409003}.

\end{thebibliography}
\end{document}